# Discovery and Application of the Two-Electron Quantum Theory of Glass States

Jia-Lin Wu

**AFFILIATIONS**

吴嘉麟 (Wu Jialin), College of Material Science and Engineering, Donghua University, No. 2999, Renmin North Road, Songjiang District, Shanghai 201620, China.

Email: jlau@dhu.edu.cn

**AUTHOR CONTRIBUTIONS**

The full text was conceptualized, and writing, images, reviews, and proofreading were performed by Wu.

**ETHICS STATEMENT**

This study does not include studies of human data or tissues or animals

**FUNDING S STATEMENT**

This theoretical study was completed after the author's retirement, and was not funded by a grant.

**ABSTRACT**

The glass state problem stems from the failure described in terms of one-electron theory or atoms (molecules) as independent particles. In 2005, de Gennes proposed that the way to explain the glass transition in simple terms was to construct the cluster model of molecules in contact with all existing glass models and to refine the picture of the mean-field hard-sphere molecules (HSMs) in contact with each other. In the process of refining this picture, we discovered the two-electron quantum theory derived from the second solution of de Gennes $n$ = 0, where the clustered contact of the two HSMs along the $z$-axis is the sequential emergence of the 16 $z$-direction interface excited quantum states of their coupled electron pair, the two HSMs suddenly overlap by 0.27% to form a magic-interface two-dimensional vector. The two coupled electron orbitals synchronously escaped the two HSMs 16 times, tangent to the magic interface 16 times, and 16 parallel repulsive electron pairs with an interval of 5.9987°, which is a clustered boson interaction between the two HSMs. This is the common origin of boson peaks in the glass state and electron pairing in the high-temperature superconductivity. Therefore, the collective behavior of electrons in the two-electron theory can unify the glass transition and the high-temperature superconducting transition. This paper is not only a complete theoretical statement on glass transition, but also a new interpretation of the theory of high-temperature superconductivity, which provides a new theoretical perspective in the search for room-temperature superconducting materials.

## I. INTRODUCTION

"Philip W. Anderson's 1995 statement that 'the deepest and most interesting unsolved problem in solid state theory is probably the theory of the nature of glass and the glass transition' still stand"1. "What is the nature of the glassy state? "2 "What is the pairing mechanism behind high-temperature superconductivity?"2 In this study, we focused on finding a theoretical framework to describe the "collective behavior of electrons" to uniformly explain the glass transition and electron pairing in high-temperature superconductivity, In the solid state, "the Mott transition illuminates a regime in which one (single)-electron or independent particle of solids fails" [3]. The Mott transition implies the existence of a two-electron theory, in which the repulsion of two electrons at the two-molecular interface increases the potential energy of the system. However, thus far, solid-state physics has been blank in terms of two-electron theory. The aim of this study is to refine the cluster contact picture of HSMs and fill the gap in two-electron theory. In Mott transition, electrons form lattices to avoid meeting each other and minimize the total dominant potential energy[3]. Inspired by the Mott transition, glass transition is also a failure of the one-electron approach. The Mott transition excludes all double-electron repulsive states at the interface; however; in the glass transition, it is the search for all possible two-electron repulsive states in the *z*-direction of the coupled electron pairs (CEPs) of all adjacent molecules in the largest two-dimensional (2D) cluster (soft matrix; see Section **II.B.*3***). Thus, the total potential energy of the soft matrix reaches a maximum value ($k_B T_g°$) in the *z*-direction, inducing all the molecules in the soft matrix to *collectively jump* a tiny step along the *z*-direction within the covalent bond scale (Section **III.E.*1***). Our approach is to first establish nine 2D clusters from small to large along the *z*-axis of the central $a_0$-HSM according to the inverse-cascade mode of fluid dynamics, and then search for and identify the theory that can derive all possible orientation states of the CEP at each interface along the *z*-axis, which has never appeared in the existing one-electron mathematical-physical systems.

Traditional glass state studies are based on one-electron theory, where mathematical and physical equations are established based on certain properties of glass and glass transitions. Subsequently, the energy minimum or extreme value of a parameter is determined and compared with the experimental results. There are now more than a dozen such models, including free volume, cage, trap, kinetic, and thermodynamic models; replica symmetry; potential energy landscapes and heterogeneity; mode-coupling and random first-order transitions; Kauzmann temperature and percolation transitions; fast–slow relaxation; pinning; jamming; and boson peaks, **e**ach of which can be explained online, each of which has a reason for its existence, and each describes a property of the glassy state. However, none of these models alone can reveal the nature of the glassy state. Note that there is no clustering model for molecules in the existing models; therefore, no pattern of the collective jumping of molecules has been found thus far**.** In 2002, de Gennes highlighted the need to establish a "cluster model of molecules"[4] that engages with a variety of existing schools of thought (models). In 2005, he further noted that the way to explain glass transition in simple terms was to refine the picture of the mutual cluster contact between mean-field HSMs[5]. This picture touches on the nature of the glass state: What is the interaction between two neighboring HSMs in the mean-field? How do they cluster? The unexpected result of this study is that the two HSMs must undergo a critical state mutation in the HSM model, which suddenly overlaps by 0.27% to form a magic interface, and at the magic interface, they complete 16 consecutive cluster-contact interactions along 16 pairs of coupled cluster-contact angle lines (a new form of eigenvectors of disordered systems) on two orthogonal diagonal planes, 16 new pairs of synchronously coupled two-electron orbitals (*16 pairs of second-order* $\delta_{\lambda\lambda^*}$ *orbital magnetic moments*) tangent to the magic interface 16 times, yielding 16 pairs of non-zero eigenvalues, which are the 16 z-axial repulsive electron pairs of the CEP that appear sequentially at the magic interface.

Several researchers have also proposed similar methods. In 1985, for example, one scholar noted that the crucial this endeavor is "to deeper understanding of *the systematics of bonding* in condensed matter within a framework going considerable beyond the current picture"[6]. In 1975, the spin-glass image envisaged by Edwards and Anderson constructed a simple model and "incorporated the two physical components of geometric frustration and quenched disorder into the lattice Hamiltonian and the Ising model" [7]. However, the problem of quenched disorder is similar to the glass state problem, and remains an open question. Combined with de Gennes' comments, the shortcoming of the existing spin glass theory is "no-neighborhood effect”[5] Here, de Gennes makes it clear that the direction of research in spin glass theory is to find the interaction that occurs only between two adjacent HSMs. This is reminiscent of the $n$ = 0[8] second-order $\delta_{\lambda\lambda^*}$ vector, discovered 50 years ago by de Gennes. "This so-called '$n$ = 0 theorem turned polymer physics upside-down, allowing theories of phase transitions to be applied to polymers and earning him the Nobel prize in 1991."[9] In looking for the clustered interaction between two adjacent HSMs, we found the second solution of $n$ = 0 theory (Section **III.A**).

## II. METHORDS

Methods in this chapter refer to the theory of how to find glass state.

### A. The germination of the two-electron theory

The germination of the two-electron theory came from the study of polyester melt super-high-speed spinning theory. The polyester melt super-high-speed spinning line links the three challenges of anomalous melt viscosity, 30000%-fold super-tensile fluid mechanics, and orientation glass transition along the $z$-axis at a distance of 3 m and within a few milliseconds, indicating that these three challenges involve the same inverse-cascade model of five HSMs/five clusters/five local fields and the collective jump mode of molecules (**Supplementary Materials 1**, **SM 1**). Thus, the glass state theory can be explored by solving the mystery of the anomalous viscosity of the 3.4 power law of entangled polymer melts[12]. The approach to explore the anomalous viscosity of melts in this study is also distinctive, starting with an accurate estimate (*this step made the theoretical exploration less detoured*) of the universal expression for entangled polymer melts as $\eta \sim N \exp [9 (1-T_g /T_m)]$, which is in good agreement with all known experimental data for flexible and nonflexible polymers. As this is an exponential relationship, it is highly sensitive to the assumed theory, and only a good theory can prove this equation. Thus, a theoretical derivation of the expression for this anomalous viscosity makes it possible to determine the theory of the glassy states. This expression contains two temperature-independent ordering energies, $k_B T_g^{\circ}$ and $k_B T_m^{\circ}$, where $T_g^{\circ} / T_m^{\circ} = T_g / T_m$ [Equation (12)] holds for all molecular systems. This is one of the most important advances in understanding ideal disordered systems. The inability of the one-electron theory to derive these two ordering energies also adds confidence to the search for two-electron theory. The three articles[10, 11, 12] link three challenges, and five corollaries provide a dynamic ordered structure of an ideal disordered system. **(i)** The critical molecular weight 200-HSM of an entangled polymer chain consisting of can only be obtained using the 5-HSM / 5-cluster / 5-local-field and inverse-cascade in fluid mechanics with the ninth 2D cluster interruption. **(ii)** To achieve an inverse-cascade, the nine 2D clusters $V_i$ ($a_0$), $i$ = 0, 1, 2, 3,..., 8, centered on $a_0$-HSM, from small to large, must be 2D vectors, i.e., each molecule has a 2D vector in the form of an HSM cubic lattice (HSCL) along the $z$-axis, which is also defined as the molecular interface excitation vector (MIEV); adjacent two MIEVs are in opposite directions, thereby satisfying stable Ising state conditions. **Table 1** shows the data of these nine clusters. **(iii)** Thus, in the $x$–$y$ projective plane, the $V_0$ ($a_0$) cluster is a +$z$-axis vector formed by four consecutive arrows around $a_0$, denoted by the +z-$V_0$ ($a_0$) loop. It is clear that the two adjacent closed loops are in opposite directions and in the Ising state. **(iv)** In the molten state, a chain of length $N$ is labelled $N$-chain, and the movements of the three component chains of $N$-chain ($N_z$, $N_x$, and $N_y$) are both independent and entangled with each other. *Independent*: refers to the number of degrees of freedom (DoF). $N^*$ required for the complete free diffusion of an $N$-chain is not proportional to $N$ and must be $N^* = N_z^* \cdot N_x^* \cdot N_y^*$, where $N_z^*$ ($N_x^*$and $N_y^*$) is the number of DoFs required for a $z$-component chain to spread freely in $z$-space, which exceeds the Debye model describing the number of DoFs in crystals. This also suggests that, *in each local area of the glass transition, only component chains in a certain direction (e.g., $N_z$ chains) are excited to jump along the z-axis in a tiny step $n_z \leq 0.036$* (within 0.1 of a vibration amplitude of the covalent bond). *Entanglement*: According to de Gennes, the (largest) clusters move rather than the molecules[6], and the first chain-unit $a_0$-HSM in the $N_z$-chain can only move along the +$z$-axial direction by establishing the +$z$-$V_8$ ($a_0$)-soft matrix (which means that the 2D $V_8$ cluster centered on $a_0$ in the $x-y$ projection plane contains 200 $z$-component chain units-HSMs, which lie on 200 other $z$-component chains). However, the energy required to build the +$z$-$V_8$ ($a_0$)-soft matrix must satisfy the condition that the $N$ +$z$-$V_8$ ($a_j$)-soft matrices in the $N_z$-chain can walk one after the other, $j = 0, 1, 2, 3\ldots N\text{–}1$, and each soft matrix jumps sequentially with the same step size, although the $j$-th soft matrix has not yet appeared in the observation time, which is the solitary-wave interface excited mode of $N_z$-chain walking12. Furthermore, each $a_j$-HSM in the $N$-chain in the melt can successively construct three soft matrices in three of the six directions along the ±$z$-, ±$x$-, and ±$y$-axes, for example, $z$-$V_8$ ($a_0$), $x$-$V_8$ ($a_0$), and $y$-$V_8$ ($a_0$). Characteristically, if $a_0$-HSM selects the +$z$-$V_8$ ($a_0$)-soft matrix (**SM1, Figure S1a**), all $a_j$-HSMs in the $z$-component chain can only select the same +$z$-$V_8$ ($a_j$) soft matrix as+$z$-$V_8$ ($a_0$). **(v)** *The soft matrix travels $n_z$ steps in disappearance*, and *its disappearance is due to the sequential appearance of four adjacent soft matrices*. The five inferences derived from the theoretical proof of the anomalous viscosity expression of the entangled polymer melt sketched a rough picture of the glass transition. However, for the inference to hold, it must be shown that the interface between two neighboring HSMs is a 2D vector, which is both the hardest thing to do theoretically and narrows the search for the new theory to the overlapping interfaces of the two HSMs, allowing us to eventually find the second solution for $n$ = 0 (Section **III.A**).

**TABLE 1. Clustering data of molecules** (Based on the supplementary materials SM 3)

| $V_i$ | Number $N_i$ of HSMs in cluster | Number of C-boson on $V_i$-loop | Number of C-boson in $V_i$-cluster | Number of HSCLs in $V_i$-loop | Number **of** Jamming particles | Number of +$z$-HSCLs | Number of –$z$- HSCLs | Relative density of cluster* |
|---|---|---|---|---|---|---|---|---|
| $V_0$ | 5 | 4 | 4 | 1 | 0 | 1 | 0 | 1 /5 = 0.2 |
| $V_1$ | 17 | 12 | 16 | 5 | 8 | 1 | 4 | 5/17 ≈ 0.29 |
| $V_2$ | 33 | 20 | 36 | 13 | 16 | 9 | 4 | 13/33 ≈ 0.39 |
| $V_3$ | 53 | 28 | 64 | 25 | 24 | 9 | 16 | 25/53 ≈ 0.47 |
| $V_4$ | 77 | 36 | 100 | 41 | 32 | 25 | 16 | 41/77 ≈ 0.53 |
| $V_5$ | 105 | 44 | 144 | 61 | 40 | 25 | 36 | 61/10 ≈ 0.58 |
| $V_6$ | 137 | 52 | 196 | 85 | 48 | 49 | 36 | 85 /137 ≈ 0.62 |
| $V_7$ | 173 | 60 | 256 | 113 | 60 | 49 | 64 | 113/173 ≈ 0.65 |
| $V_8$-soft matrix | 196 + 4 | 56 + 4 | 312 + 8 | 136 + 5 | 60 | 60 +16 | 60 | 136/200 ≈ 0.68** |

* The number of HSCL contained in $N_i$ HSMs in a cluster is defined as the relative density of $V_i$.

** The relative density of the soft matrix, 0.68, does not account for the contribution of the 5 vacancy-volumes at the center of the soft matrix because in the mean-field, the four adjacent cavity volumes of $a_0$ belong to the four adjacent soft matrices of $a_0$-soft matrix.

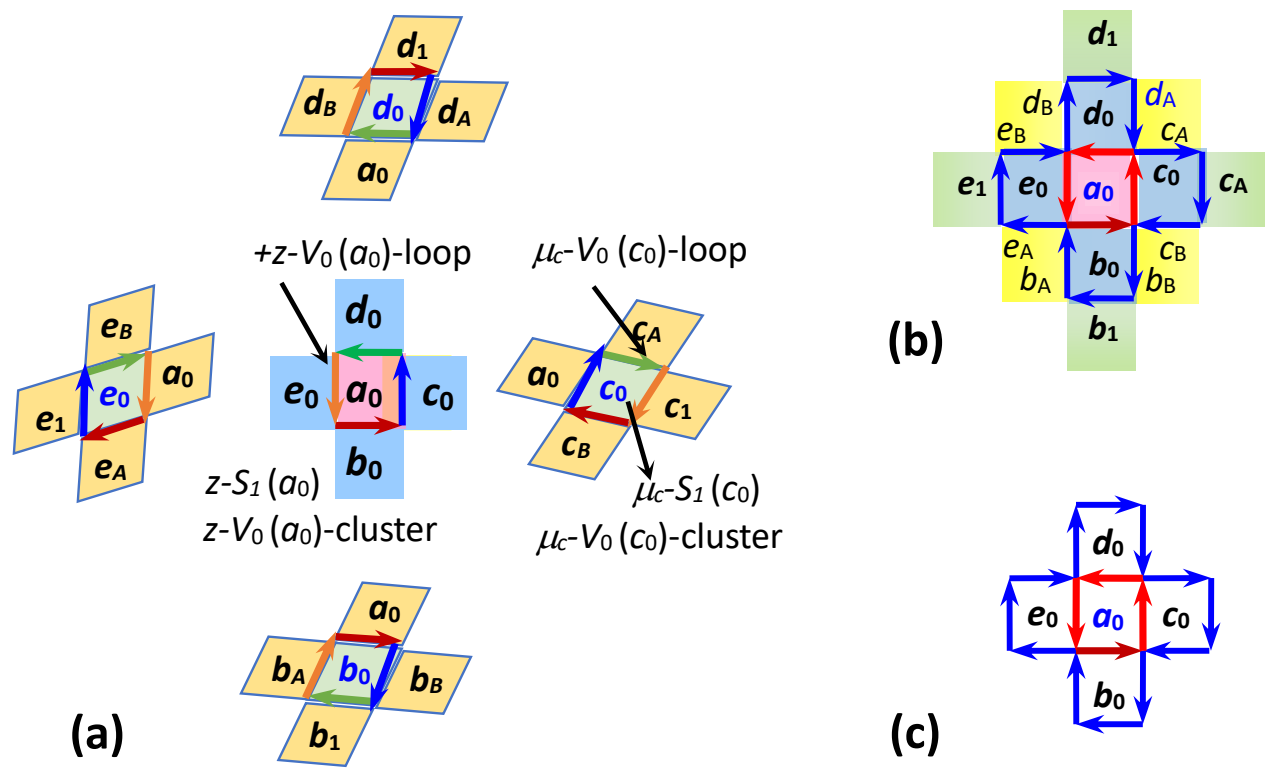

**Figure 1.** Each arrow represents a magic interface in Section **III.D.*1*. a.** $V_0$ ($a_0$) cluster, $V_{0}$-($a_0$) loop, MIEV $S_1$($a_0$) of $a_0$, and four $V_0$-clusters of four HSMs adjacent to $a_0$. **b.** Four adjacent $V_0$-clusters are sequentially projected onto the $z$-axis and overlapped with the $+z$-$V_0$ ($a_0$) to generate a $-z$-$V_1$ ($a_0$) cluster with 8 jamming particles. The 12 successive blue arrows indicate the $-z$-$V_1$ ($a_0$) loop, and the center four successive red arrows indicate the MIEV $+z$-$S_2$ ($a_0$). **c.** It is agreed to include an HSM on the outside of each arrow to simplify the representation of$-z$-$V_1$ ($a_0$) cluster and $-z$-$V_1$ ($a_0$) loop.

## B. Discovery of eigenvectors and eigenvalues for disordered systems

### *1. Complementarity of one-electron and two-electron theories*

The nature of the glassy state is found in symmetry breaking in ideal disordered systems (e.g., flexible polymer chains), which has been a topic of interest for scholars- for example, Gil Refael proposed that "the essence of universality in disordered quantum systems: the low-energy physical properties are independent of the disorder distribution"13, and that "fermions are confined to pairs of sites," and that "the pair formation is scale invariant." This view is similar to that of the two-electron theory. The difference is that the "two fermions formed in pairs are located at lower-energy sites in the local phase transition"13, whereas the two positively charged particles (PCPs) of two electrons in the two-electron theory are located at two discrete 1/16 equipotential points, **Figure 2a**; these two electrons are interface excited states in the Mott transition, as shown in Figure 5.12 in Ref.3**.** Recent use of "machine learning"14,15, as well as the use of state-of-the-art electron microscopy techniques in combination with machine learning, attempts have been made to "study ordered parameters and dynamic processes in disordered systems"16. However, the processing of data by this powerful tool is still based on one-electron theory, Noting interesting phenomena based on exotic metals, spin glasses, and superconductivity in recent years, terms such as "quantum critical point" and "many-body quantum state"17, 18, 19 also seem to be new explorations that have been made with the failure of the one-electron approach. However, these are all studies of a particular material, and the understanding of whether there are ordered quantities in an ideal disordered system is basically "negative"13. Ideal disordered systems "do not have clean counterparts for various interesting phenomena17. Whether it is a "quantum critical point transition18,19 " or a "many-body quantum state17", they is still a product of the one-electron theory. These terms actually point to a two-electron theory, in which they are very simple**:** the quantum critical point near the absolute temperature corresponds to the two-electron approach, in which the system enters the glass state, and the many-body quantum states correspond to the CEP 16 $z$-axial interface excited states in the 2D $z$-$V_0$ cluster.

### *2. Symmetry breaking in ideal disordered systems.*

In 2021, based on the cluster model envisioned by de Gennes4, we found that there are ordered physical quantities or symmetry breaks in the cages of HSMs in the glass state of an ideal disordered system. **(i)** The orientation of the cubic lattice cage of all atoms (molecules) in each local region is along the same direction (e.g., $z$-axis) to comply with the corollary presented in Section **II.A(iii)**, that is, the orientation of the cubic lattice with a constant side length of 2Δd, breaks the isotropy of the spherical cage. **(ii)** The PCP makes a closed-loop jump along only $k_\lambda$ discrete equipotential fixed points, breaking the symmetry of the PCP along 4-diagonal closed-loop parallel transport. **(iii)** The clustering of each molecule with adjacent molecules occurs only at $k_\lambda$ discrete equipotential fixed points in 360° space around the z-axis, resulting in only $k_\lambda$ delta-clustered angle-line vectors for the central molecule, breaking the symmetry of the angle lines along any angular direction, which is similar to the angular quantum numbers in quantum mechanics.

### **3.** *Discovery of eigenvectors and eigenvalues in ideally disordered systems***.**

The intuitive model of this study, **SM3**, obtained the largest 2D $V_8$-cluster containing a cavity left by the central HSM being crowded out of the center; however, a theoretical basis for the arrangement of the two-electronic orientation at the interface needs to be found. The term soft matrix, coined by de Genned in his cluster contact picture, refers to a substrate that contains a cavity, is low-density, easily flowable, and fills hard-sphere gaps**.** Comparison with $V_8$ cluster suggests that the mobile substrate filling the HSM gaps is the electronic orbital that escapes these HSMs and is tangential to the overlapping interfaces. As a result, the "cross-shaped parallel transport" of CEP interface excited state in the early intuitive model should be corrected to the "cross-shaped parallel jump transport" The "HSM interface excited spin" of $n = 0$ shared by monoatomic metallic glass and polymer glass is discovered. Here, spin refers only to the spin in the sense of the spin component number $n = 0$ or the spin in the spin glass, which is a pseudospin in all glass states except in the case of superconductivity. In order to avoid misunderstanding, this paper changes the name of "HSM interface excited spin" to "HSM interface excitation vector" (MIEV).

## C. Advances and shortcomings of cluster model in the previous article

In the previous article [20], the cluster model of molecules has made important progress in three aspects. **(i)** Each molecule in a disordered system, in addition to the HSM form in the mean-field, can sequentially appear one to three 2D vectors in the form of a dynamic cubic lattice along the *z*-, *x*-, and *y*-axes as temperature increases throughout the solid-liquid transition. **(ii)** In a disordered system, in addition to the random thermal vibration in the 2Δd cubic lattice cage, the PCP of each HSM can also perform one to nine closed-loop jumps along the $k_\lambda$ discrete equipotential points on the 4-diagonal of the cage.. The time taken for the PCP of the HSM to complete the *i*-th closed loop in the 4-diagonal of the cage is the *i*-th relaxation time in the system. **(iii)** In the glass transition, there is only a density increase caused by the overlap of the two HSMs, and the density of the clusters in the inverse cascade increases with the scale, as shown in **Table 1.** Therefore, it is more appropriate to replace the density discussion of the mode-coupling theory with the cluster model because de Gennes pointed out that the mode-coupling theory ignores geometric setbacks[5]. Table 1 lists the data changes in the soft matrix owing to geometric frustration.

In the previous article[20], we simply expressed the discrete equipotential point of the PCP of $a_0$-HSM along the 4-diagonal of the cage as $k_\lambda$ (= $k_\alpha + k_\beta + k_\gamma + k_\delta$), and one of the contents of this paper is to find the values of $k_\lambda$ and the number of interface excited states of four CEPs along the *z*-axis on the four interfaces, $k_\alpha$, $k_\beta$, $k_\gamma$ and $k_\delta$. When solving the value of $k_\lambda$, it is found that there are two errors in the previous article[20]. One is in the $\beta$--space of $a_0$ and $c_0$, where the electron does not form a spiral precession in the $k_\beta$ electron orbitals, but the $k_\beta$ (=16) simple electron orbital quantum states emerge along the $k_\beta$ angle lines**.** The other is to correct the "$k_\beta$ collision overlap" of $a_0$ and $c_0$ in $\beta$-space to "at the overlapping magic interface, two PCPs of $a_0$ and $c_0$ jump synchronously 15 steps along two orthogonal diagonals in tiny steps of 0.01". Thus, the 15-step sequential jump yields a clustered boson (C-boson) and a two-electron trap state of 0.019° (Section **III.D.*1*).**

## D. An invariant in disordered systems

In the glass model of the random first-order theory"[21], the Lindemann ratio $d_L$ is the maximum amplitude of a thermally vibrating molecule**.** However, the original Lindemann $d_L$ (about 0.1) failed in the solid-liquid transition[22]. In the two-electron theory, we do not use Lindemann ratios, but define the invariant that transform the small cluster scale into a large cluster scale in the inverse cascade of disordered systems as $d_L$*. $d_L$*= 0.10464895…, which still retains the maximum amplitude of the molecule, but has more properties. The discovery of the invariant $d_L$* is a critical step in the creation of the two-electron theory, It is a universal constant in disordered systems. $d_L$* is independent of the melting transition, and the system enters the melting transition phase when the number density of the largest clusters (soft matrices) in the system reaches a critical value of 5, Section **III.F.*2*.**

## III. **TWO- ELECTRON THEORY**

### A. The second solution of de Gennes *n* = 0

#### *1.* What is the solution of *n* = 0?

In solid state physics, "band theory is one-electron independent particle theory"[4]. De Gennes rigorously proved theoretically that when the partition function describing the electron energy level degenerates to a constant (or the Hamiltonian in the quantum mechanical equation degenerates into the soft matrix lattice Hamiltonian of a certain material), the new system can be described by the second-order delta vector theory with spin component numbers $n = 0$[8]. However, the difficulty in applying the $n = 0$ theory or $n = 0$ solution is that there is no pattern to follow in finding the specific geometry in which $n = 0$ occurs for each new system. In other words, different physicochemical systems have different geometries of $n = 0$. The first solution of $n = 0$, given by de Gennes, is the self-avoiding random walk of a macromolecular chain in a 2D lattice[8]. In searching for and confirming which theory can derive the theory of "electron pairing" at the overlapping interface of two HSMs, it turns out that this is precisely the second solution of $n = 0$ found in this paper.

#### *2. Deflection of the eigenvector and parallel jump-transport of the eigenvalue*

The five-HSM clusters are closed-loop hops of $a_0$-PCP along the 4-diagonals of the cage in the *z*-direction. When $a_0$-PCP jumps along point 10 on the cage in **Figure 2** to a point near point 11, 16 electron orbitals escaping from the HSM and tangential to the magic interface appear in the 3-4-5-6 diagonal plane, and each electron orbital forming an additional magnetic moment perpendicular to the diagonal plane around $a_0$-PCP, which is equivalent to the direction of $a_0$-PCP perpendicular to the diagonal plane. Here, with two adjacent magnetic moments spaced 5.9987° apart from point 10 to point 11, the 16 additional magnetic moments of $a_0$-PCP rotate by (90 – 0.019)° around the *z*-axis. However, the eigenvalue of the 16 pairs of coupled electron orbitals tangent to the magic interface is a parallel jump transmission of 15 steps along the *z* axis.

#### 3. *Geometric conditions for the second solution of n* = 0

For example, see **Figures 2 and 4** for two adjacent atoms (molecules), $a_0$ and $c_0$. **(i)** The two adjacent *z*-component HSMs $a_0$ and $c_0$ suddenly overlap by 0.27% to form a magic interface oriented along the *z*-axis, and the $a_0$-PCP and $c_0$ -PCP of $a_0$ and $c_0$ are located at points 10 and $10^*$ in **Figure 2.** The two electrons form the first pair of orthogonal electron orbitals along the first pair of coupled diagonals ($\lambda_\beta$ = 1st and $\lambda_{\beta^*}$ = 1st) synchronously tangent to the magic interface at points 2 and $2^*$. **(ii)** Two diagonals 10-11 and $10^*$-$11^*$ on the two cages are orthogonal, which are located on the two orthogonal diagonal planes and parallel to the orthogonal diagonals 6–3 and $6^*$–$3^*$. **(iii)** From points 10 to 11 (points $10^*$ to $11^*$), there are 16 pairs of coupled equipotential points and 16 pairs of cluster contact angle lines with a constant spacing of 5.9987°. The overlap of the two HSMs causes symmetry breaking of the HSM, resulting in the synchronous escape of the two electronic orbitals of the two HSMs along the coupled $\lambda$–$\lambda^*$ angle lines, which converts the point $\theta_\lambda$ (the red dot in **Figures 4** and **5**, the point at which the two electrons of the two orbitals meet, which is not allowed in the one-electron theory) and $\theta_{\lambda^*}$ into two tangent electrons $p_\lambda$ and $p_{\lambda^*}$ tangent to the *z*-axis magic interface, satisfying

$$\delta_\lambda(\theta_\lambda) \cdot \delta_{\lambda^*}(\theta_{\lambda^*}) = p_\lambda p_\lambda \cdot \delta_{\lambda\lambda^*} \qquad (1)$$

Here, $\delta_\lambda(\theta_\lambda)$·and $\delta_{\lambda^*}(\theta_{\lambda^*})$ are the two angle lines of the cluster contact points in one-electron theory, respectively. The second-order eigenvector consists of the two coupled angle lines $\delta_\lambda$–$\delta_{\lambda^*}$, the "matrix element" acting on the second-order eigenvector is the synchronous anti-symmetrically coupled (SASC) two-electron orbitals that escape from the two HSMs and are tangent to the magic interface, and the $z$-axial $p_{\lambda}$–$p_{\lambda^*}$ repulsive electron pair as an eigenvalue, which is a simple two-electron quantum state.

### B. Clustering diagram of two adjacent atoms in a monoatomic metallic glass.

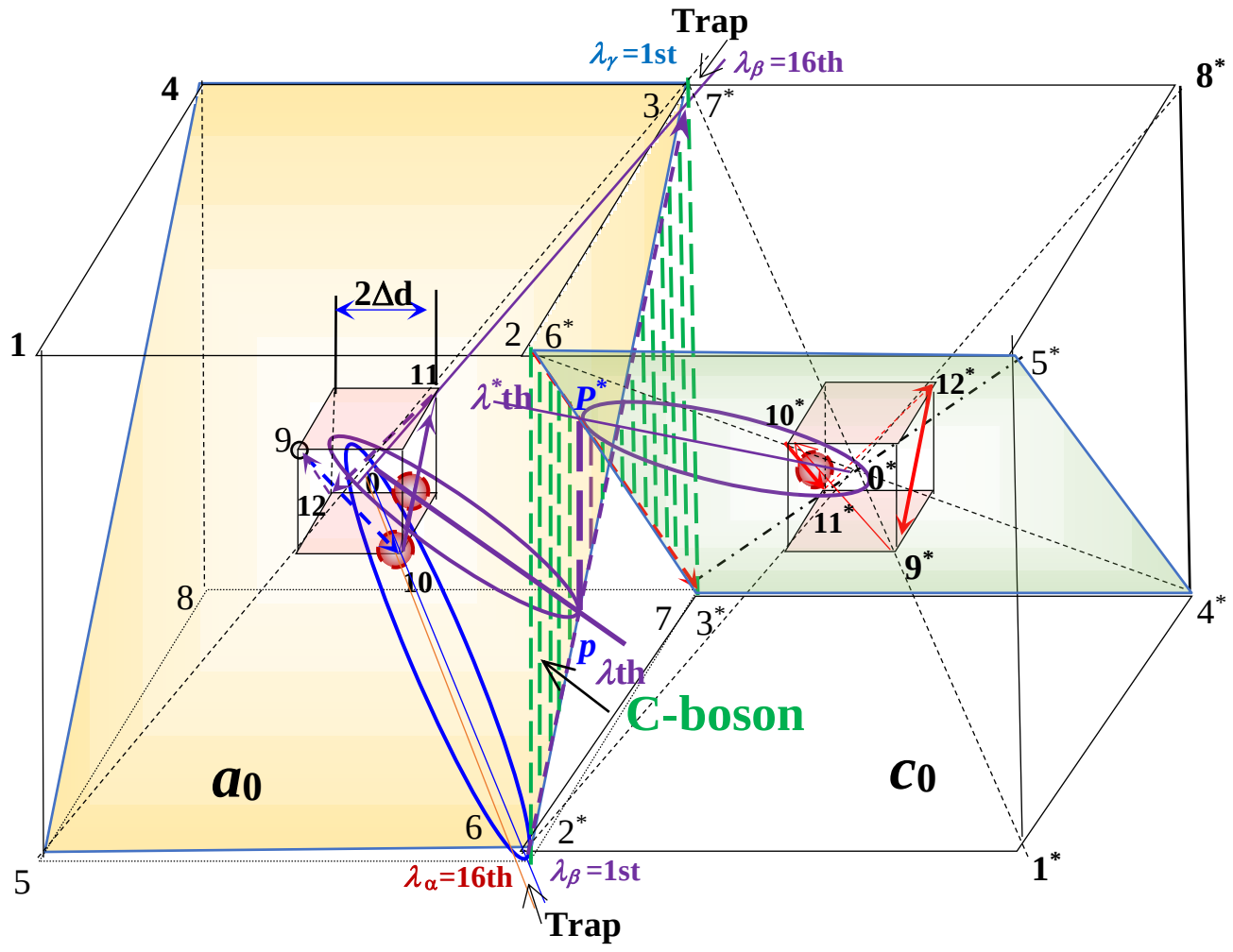

**Figure 2.** Dynamic clustering diagram of SASC of adjacent atoms .$a_0$ and $c_0$. 9 →10 →11→12→ 9 is the trajectory of the $a_0$-PCP of the $a_0$ atom making a 4-diagonal closed-loop jump along its 2Δd cubic-lattice 4×16 1/16 equipotential points. When $a_0$-PCP and $c_0$-PCP synchronously transition to the coupled $\lambda$–$\lambda^*$ angle lines the two coupled $\lambda$–$\lambda^*$ electron orbitals are tangent to the points $p$ and $p^*$ at the magic interface 2-3-3*-2*. The 16 eigenvalues that appear sequentially from 2-2* to 3-3* are 16 repulsive electron pairs, plus 16 additional pairs of electron orbital magnetic moments, which form a C-boson (2D vectors) of the $a_0$ and $c_0$ atoms. After 15 steps, the $a_0$ atom falls into the trap state angle of about 0.019° between the angle line $\lambda_\beta$ =16th and the angle line $\lambda_\gamma$ =1st,

Compared to the approach of the quantum critical point and many-body quantum state, the approach of single-atom metallic glass in the cluster model is simple, clear, and theoretically rigorous (**Figure 2**. The cluster model considers the critical state mutation of the hard-sphere model of two adjacent atoms in the mean-field. Each metal atom, in addition to having quantum mechanical morphology, also has a hard sphere $\sigma$ of L– J potential, measured with dimensionless potential well energy $|\varepsilon_0|$=1, $\sigma$ =**1.** In the overlapping $\beta$-space between the $a_0$-atom and the $c_0$--atom in the mean field, along the coupled two $\lambda$–$\lambda^*$angle lines (eigenvector), two SASC electron orbitals suddenly appear along the coupled two $z$-axial angle lines (eigenvector) and tangent to the points $p$ and $p^*$ at the overlapping interface. The 16 eigenvectors of the 16 pairs of coupling angle lines generated 16 repulsive electron pairs parallel to the $z$-axis of the $a_0$–$c_0$ CEP at the overlapping interface between the two hard spheres of $a_0$-atom and the $c_0$-atom. The 4 ×16 repulsive electron pairs (eigenvalues) parallel to the z-axis in the low-temperature $V_0$-cluster may correspond to the so-called "quantum Griffiths phase of the electron-nematic quantum phase transition"[23]

## C. Cluster model in one-electron theory

### *1. The contact angle line in clustering*

To solve $k_\lambda$ we must revisit the derivation of the five-HSM neighborhood effect $\chi$i-potential hidden in the L– J potential, and revisit the nine long-range fast-acting (relaxation time $\tau_i$) L– J potentials established in the intuitive model[11, 20].

$$f_i\,(\sigma_i/q_i) = 4[(\sigma_i/q_i)^{12} - (\sigma_i/q_i)^6] \tag{2}$$

Equation (2) contains nine short-range and slow-acting (relaxation time $\tau_{i+1}$) neighborhood effect potentials

$$\chi_i = (\sigma_i / q_i)^6 \qquad (3)$$

As can be seen from $\chi_i = 1$, the same as the "unit 1" for measuring Δd-, the "unit 1" for measuring $\chi_i$ also originates from the dimensionless potential well energy (Take $|\varepsilon_0(\tau_i)|=1$) for all hard repulsive hard spheres $\sigma_i$. The physical meaning of $\chi_i$ is as follows: In addition to the $V_i(a_0)$ loop interacting with countless $V_i$-loops in different directions in 3D space to generate the L– J potential $f_i$, $V_i(a_0)$ loop can also interact with $V_i(b_0)$, $V_i(c_0)$ loop, $V_i(d_0)$, and $V_i(e_0)$ loops to generate a $V_{i+1}(a_0)$ cluster. The sudden appearance of a single $V_{i+1}(a_0)$loop (relaxation time $\tau_{i+1}$) corresponds only to *an additional delta potential* $\Delta\chi_{i+1}$ in the neighborhood effect potential $\chi_i$ of the system. That is, the $a_0$-field has two interaction potentials, fast ($\tau_i$) and slow ($\tau_{i+1}$), and the relationship between the two is:

$$f_i(\chi_i) = -4\chi_i(1-\chi_i) \qquad (4)$$

The balance of the two potential fluctuations is

$$\Delta f_i(\chi_i) = \Delta\chi_{i+1} = \Delta\chi_i(\partial f_i / \partial\chi_i) \qquad (5)$$

Equation (5) indicates that the additional potential $\Delta\chi_{i+1}$ superimposed on the $z$-axis always cancels out the incremental $\Delta f_I(\chi_i)$ of the $z$-axis $f_i(\chi_i)$ potential along the z axis. This is because the direction of the vector $V_{i+1}$-loop is always opposite to that of $Vi$-loop. The stability conditions in equation (5), $\Delta\chi_{i+1}$ must not exceed $\Delta\chi_i$: **24**

$$|\Delta\chi_{i+1}/\Delta\chi_i| \le 1 \qquad (6)$$

From equations (2) to (6) and

$$\partial f_i / \partial\chi_i = \pm 1 \qquad (6\text{-}1)$$

$$q_{i,R} = q_{i+1,L}, \qquad (6\text{-}2)$$

*we* obtained nine fixed points of nine L– J potentials: $f_c = 1/16\ \varepsilon_0(\tau_i)$ at the nine cluster positions on the $q$-axis in Figure 1 in ref**20**, and $\chi_{\min} = 3/8 = (\sigma_i / q_{i,R})^6$, $\chi_{\max} = 5/8 = (\sigma_{i+1}/q_{i+1,L})^6$. From $q_{i,R} = q_{i+1,L}$, we get a set of recursive equations: $q_{i,R} = (8/3)^{1/6}\sigma_i$ and $\sigma_{i+1} = (5/3)^{1/6}\sigma_i$ ; $q_{i+1,R} = (8/3)^{1/6}\sigma_{i+1} = (8/3)^{1/6}(5/3)^{1/6}\sigma_i$, thereupon, $\Delta q_i = q_{i+1,R} - q_{i+1,L} = q_{i+1,R} - q_{i,R} \approx 0.10464895\ \sigma_i$, and the vibration equilibrium position, $\chi_0 = 1/2$, $q_0 = 2^{1/6}\sigma \approx 1.12246\ \sigma$. A universal constant $d_L$* in a disordered system can be obtained**20**.

$$d_L^* \equiv \Delta q_i / \sigma_i = (q_{i+1,R} - q_{i+1,L}) / \sigma_i = (q_{1,R} - q_{1,L}) / \sigma_0 \equiv (q_R - q_L)/\sigma = 0.10464895\ldots \qquad (7)$$

Equation (7) indicates that the transformation of $\sigma_i$ to $\sigma_{i+1}$ boils down to the transformation of $\sigma_0$ to $\sigma_1$ because the recursive relation $q_R = q_{0,L}$, *the key clustering details from $\sigma$ to $\sigma_0$* (i.e., $\sigma_0$ to $\sigma_1$) *are missing* here. defined $d_L$* not only as a *vector* of the *irreversible maximum amplitude* of the HSM jump from the repulsive 1/16 equipotential point $q_L$ to the attracted *sharp-angled* fixed point $q_R$, as shown by the blue arrow in **Figure 3a**, but also as a *directional jump angle* $\Delta\Omega \equiv (q_R - q_L)/\sigma$ (= $d_L$*) (*the unit of measurement is dimensionless radians with an angle of 60°*) generated by two adjacent *cluster-contact angle lines* on a $2\pi$ circle with a radius of $\sigma = 1$ in the $x-y$ projection plane of $z = q_0$. Note: The physical purpose of the $2\pi\sigma$ circle for $\sigma = 1$ is to maintain the critical state of the mean-field HSM. This physical quantity appears in the molecular clustering, and a $2\pi\sigma$ closed loop around the central HSM generates a $\sigma_0$-hard-sphere. **Definition**: In 5-HSM clustering, the line that crosses a 1/16 equipotential discrete point from the average position point $0^*$ of $a_0$-HSM is the cluster contact angle line of $a_0$-HSM, and the contact between $a_0$-HSM at point $0^*$ and the adjacent $c_0$-HSM on the cluster contact angle line of $a_0$-HSM is called the cluster contact state of $a_0$-HSM. In the two-electron theory, the two SASC cluster contact angle lines of $a_0$ and $c_0$ are a second-order eigenvector in clustering of the two-HSMs in a disordered system. Thus, when $a_0$-HSM jumps along the 4×16 equipotential discrete points on the 4-diagonal of the 2Δd cubic-lattice, it carries 16 cluster contact states for each of the four adjacent HSMs, and 4×15 $\Delta\Omega$ (plus four trap states) closes a loop around the $z$-axis (closed loop of parallel jump transport around the $z$-axis in topology), completing the transition from $\sigma$ to $\sigma_0$ at the closed-loop point $q_R = q_{0,\mathrm{L}}$, as shown in **Figures 3b and 4.**

The L– J potential is considered an approximate function because its $\sigma$ and $q$ cannot be directly used to describe the data for a real atom. However, Equation (2 –7) may be strictly correct if *dimensionless energy metrics σ and q* are used to study the fixed points of the attraction-repulsion equilibrium of HSMs at different scales. A wonderful thing about the L– J potential is that when $\chi$-potential is used as a variable, it is a symmetric quadratic function in equation (4). The special balance between $\chi$-potential and the L– J potential is the slope $\partial f_i/\partial\chi_i = \pm 1$, which yields a 1/16 equipotential plane of $q = q_R$ along the $q$-axis. The detail is that the "$c_0$-HSM projects into the $z$-axial" is actually the 16 $z$-space azimuth component of $c_0$-HSM" (each abbreviated as $z$-$c_0$) completing a 15-stepa parallel jump-transport along $z$-axial in a overlapping 90° space. The "cage" of $z$-$a_0$ is a $z$-axis 2Δd cubic-lattice consisting of four 1/16 equipotential planes of four L– J potentials in the $\pm x$ and $\pm y$-axis directions in **Figure 3b**, rather than the cage in **Figure 3c.**

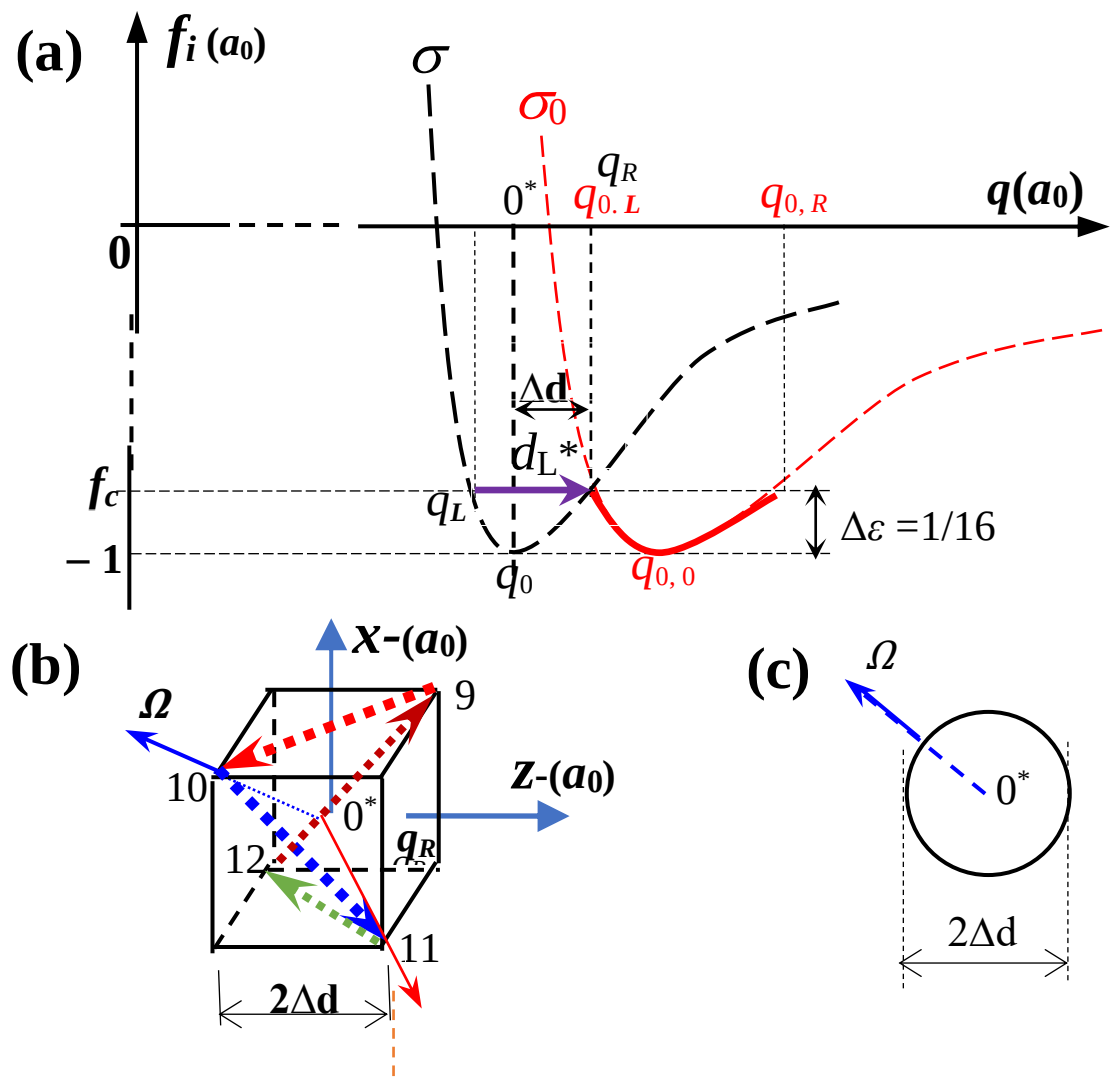

**Figure 3**. **a.** The redefined $d_L$* is both the lattice spacing (1+ $d_L$*) between two adjacent HSMs and the directional jump angle $\Delta\Omega$ of the clustered interaction when $z$-$a_0$ jumps from repulsive point $q_L$ (the point $\chi$ = 3/8) to attracted point $q_R$ (the point $\chi$ = 5/8), i.e., in $d_L^* = (q_R - q_L)/\sigma$; $d_{L^*}$ is also the largest irreversible jump spacing satisfying the 1/16 equipotential condition of L– J potential. **b.** The jumping of the centroid of $z$-$a_0$ along the 4-diagonal closed-loop of 9→10→11→ 12→ 9 is a path of 4×16 cluster-contacts of $z$-$a_0$ with $z$-$b_0$ and $z$-$c_0$ and $z$-$d_0$ and $z$-$e_0$ Denote the 0→10 angle line as $\Omega$, and represent the 90°-space between the plane (point 10 – $0^*$→ $z$-axis) and the plane (point 11– $0^*$→ $z$ –axis) as the $\beta$-space between $z$-$c_0$ and $z$-$a_0$. **c.** If it is molecular motion rather than (soft matrix) cluster, $a_0$-HSM will have a 1/16 equipotential spherical surface (or cylindrical surface), composed of L– J potentials in all $q$-axis directions in 3D (or 2D) space, neither mode meets the requirements of **II.A.(iv)**.

### *2. Key factors in the mean-field: the cage structure*

The concept of a cage in the two-electronic approach significantly simplifies glass state theory. The cage of the HSM is an ordered quantity in a disordered system. Section **II.B.*2***. It is also the simplest geometry that can unify more than a dozen existing models (Sections III.D.***3, 4,*** and ***5*** and **E** and **F**. A mention should be made here of Patrick Charbonneau et al. argued that mode-coupling theory and random first-order transition had ignored a key factor in the transition from a liquid-cooled state to a glassy state: the transition "involves self-caging, which provides an order parameter for the transition"[25]. The cage shown in **Figure 3b** is of the following importance: **(i)** Correct the geometry of the cage so that it cannot be "spherical cage"[26]. **(ii)** In addition to random thermal vibrations, the PCP of the HSM makes one to nine closed-loop jumps (equivalent to the principal quantum number in the two-electron theory) along discrete points on the 4-diagonal of the cage. **(iii)** The connection between each discrete point on the 4-diagonal and the bottom point $0^*$ is a cluster contact angle line between the central HSM and its adjacent HSM in the one-electron theory. In **Figure 3b,** the average position of $a_0$ in its 4-diagonal closed loop around the *z*-axis coincides with its vibrational equilibrium center point $0^*$, but *z*-$a_0$ obtains 1/16 of the well energy of point (point9) at the end of the closed loop. When $a_0$ completes the closed loop along the 4-diagonal line in **Figure 3b**, a 5-hard-sphere $\sigma_0$ appears at point 9. However, we cannot obtain the 2D $V_i$ ($a_0$) loop vectors and the number of $\sigma$ in the $V_i$-cluster in **Table 1,** resulting in equation (2) not being strict in the one-electron theory**.**

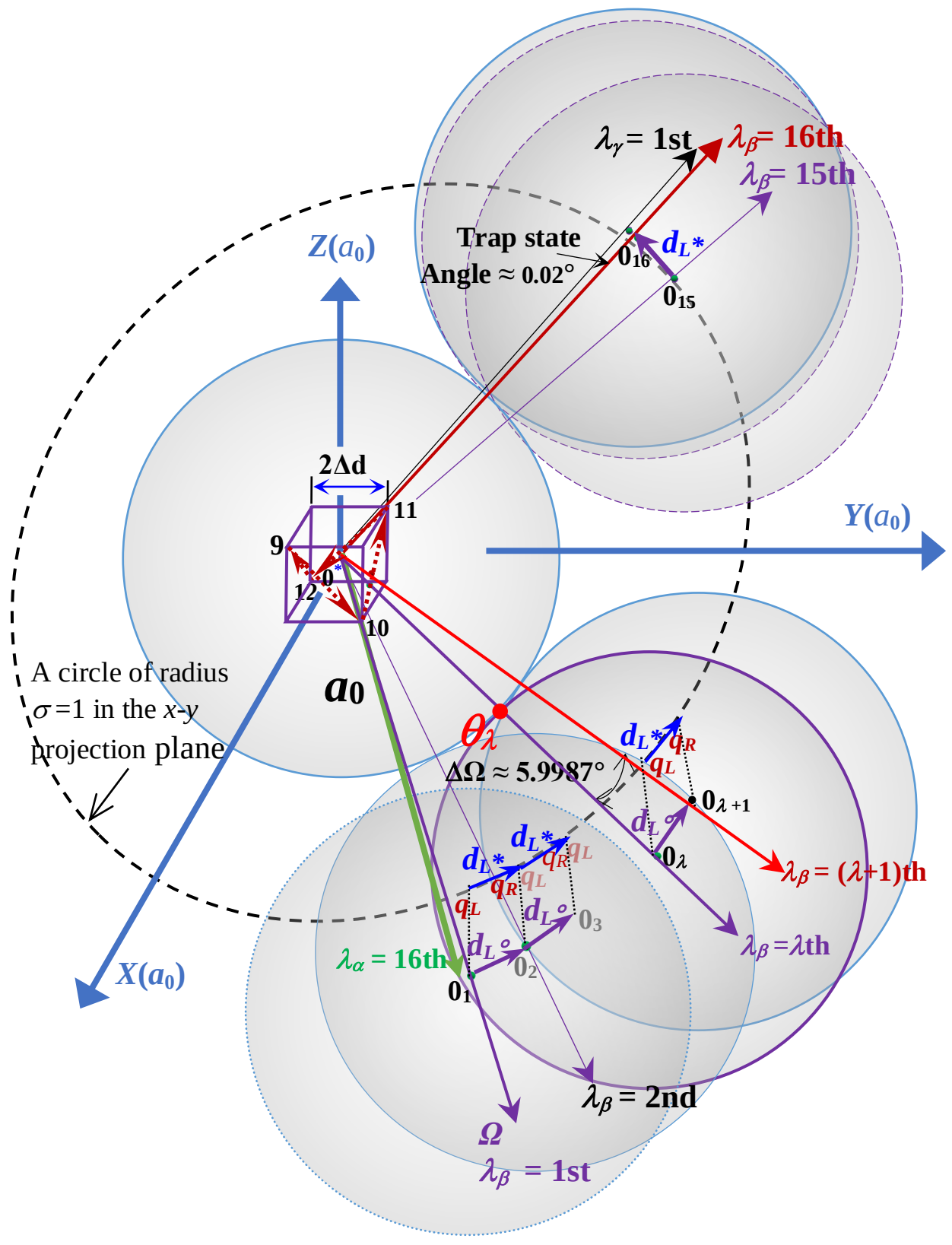

**Figure 4**. Each jump of $a_0$-HSM on the equipotential diagonal corresponds to an increase in jump angle $\Delta\Omega$ around the *z*-axis. In **Figure 3b,** $a_0$-HSM jumps 15 steps from the $\Omega$-angle line (at point 10) to the 16th angle line (at point $0_{16}$) in $\beta$-space. The 16 cluster-contact positions of *z*-$c_0$ are points $0_1$, $0_2$, $0_3$……$0_{16}$. The angle between the plane (point $0_1$–*z*-axis) and the plane (point $0_2$–*z*-axis) is the jump angle $\Delta\Omega$ of the first step of the *z*-$a_0$ jump diagonally. The two points at which the plane (point $0_1$–*z*-axis) and the plane (point $0_2$–*z*-axis) intersect with the dotted circle are the points $q_L$ and $q_R$ of *z*-$c_0$ in the cluster contact with *z*-$a_0$ during the first jump of *z*-$a_0$. The distance between points $0_1$ and $0_2$ is $d_L° = 109.5°/90°\ d_L*$. Point $0_1$ corresponds to point $q_L$ with no attraction and only repulsion in **Figure 3a**, where the left and right sides of point $0_2$ correspond to the attractive fixed point $q_R$ and the repulsive fixed point $q_L$, respectively, and so on, until the 16th point only attraction without repulsion, the HSM enters a trap state of 0.019°. We get $k_\beta = 16$ from $\Delta\Omega = 2\sin^{-1}d_L/2 \approx 5.9987°$. $\lambda_\gamma$ =1st is the first angle line between $a_0$ and $d_0$ in the 90°$\gamma$-space.

## D. Cluster model in two-electron theory

### 1. *The mutation of two molecules in clustering transcends the concept of a phase transition*.

Note that in **Figure 4**, there is a clustered contact point of two HSM on each angle line derived by the one-electron approach, which causes the two synchronous orbital electrons to meet and overlap, which is unacceptable in the one-electron theory**.** When the two SASC $z$-$a_0$ and $z$-$c_0$ [$a_0$ rotate 180°around $y = (1+ d_L^*) / 2$ axis to coincide with $z$-$c_0$] touch each other at 1/16 equipotential of their respective L – J potentials, the $z$- $a_0$ HSMs and z-$ca_0$-HSM suddenly overlap by 0.27% in the $y = (1+ d_L^*) / 2$ plane. The two SASC PCPs jump 15 steps from point 10 ($10^*$) in **Figure 5** to near point 11 ($11^*$), during which the 16 pairs of emerging electron orbitals sequentially tangent to the magic interface form 16 pairs of *z*-axially repulsive electron pairs $p$-$p^*$. This is a mutation of every two adjacent atoms (molecules) in the cluster model and not a phase transition. This is a mutation of the only permissible 16 pairs of spatial geometric angle lines (16 eigenvectors) spontaneously selected by these two atoms (molecules) in the mean-field, and the 16 eigenvalues are all oriented in the *z*-direction to become potential energy in equilibrium with the kinetic energy of the random thermal motion, called the 16 *z*-direction interface excited states of the CEP.

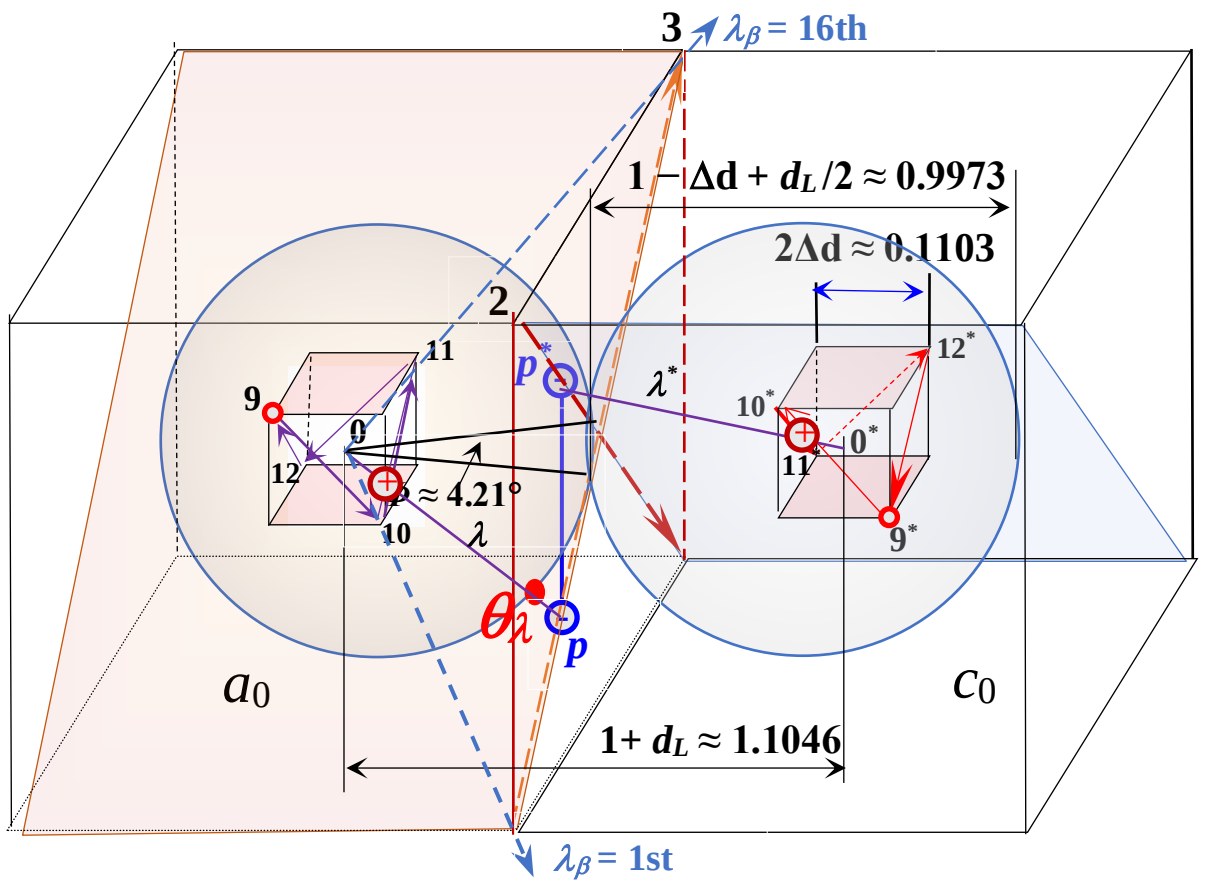

**Figure 5**. The eigenvector is two coupled $\lambda$th-$\lambda^*$th angle lines, and the overlap angle $\Phi \approx 4.21°$ is smaller than the jump angle $\Delta\Omega$ (avoiding the meeting of two electrons). From the prism line 2–$2^*$ to the prism line 3–$3^*$ is the two-dimensional dynamic magic interface, i.e., the interface excited arrow in **Figure 1,** which contains a C-boson and a two-electron trap state with an angle of 0.019°.

### 2. *New symmetry breaking at the magic interface*.

Although the 16 diagonal directions of the two hard spheres remained unchanged before and after the mutation, the left-right symmetry of the next magic interface position of the central HSM was broken. If the $z$-$a_0$ molecule in that local region chooses to be right-handed (or left-handed) to form a second magic interface, then the subsequent third and fourth magic interfaces form a $V_0$ ($a_0$) cluster on the +z axis. Left-right symmetry breaking is the most effective method for converting disordered kinetic energy to directional repulsive potential energy.

### 3. *Discovery and properties of boson*.

The 16 *z*-axial repulsive electron pairs that appear sequentially on the magic interface are caused by a new jumping pattern of two positively charged SASC particles, which jump 15 steps synchronously parallel to the *z*-axis in steps of $\sqrt{2}\ 2\Delta d/15 \approx 0.01$ along two orthogonal diagonals. The 15 synergistic jumps of the positive and negative charges within two HSMs that make the magic interface a 2D vector are the clustered boson interaction between the two HSMs, which is precisely the microscopic origin of the boson peak. Among the various schools of thought that explain the microscopic origin of boson peak, the random first-order theory considers the boson peak to be a ripplon-like domain wall motions of the glassy mosaic structure[27], very similar to 2D $V_i$ loop excitation; while the stringlets school of thought[28] regards $V_i$ loop (e.g., $V_2$ loop with 20 consecutive arrows (**SM3, Figure S2**) as one-dimensional string "vibrations", and 2D $V_2$ cluster with 33 HSMs as a local excitation.

C-bosons possess many properties that phonons do not. **(i)** *Two-molecule excitation and nine 2D cluster vector excitations*. The C-boson plus the 0.019° two-electron trap state is the 2D interface excited vector we have been looking for. Thus, the nine 2D $V_i$ clusters and 2D $V_i$ loops established in the early intuitive model of the glass state are now finally confirmed, and they are excited from small to large as the temperature changes. At the glassy state temperature $T$, the spectroscopically observed boson peak is the energy of the one-by-one emergence of the 2D C-boson in the 2D $V_i$ loop, which produces a local 2D $V_i$ cluster potential that is always in equilibrium with the disordered kinetic energy of $k_BT$. They are the "local excitation"[29] described in the one-electron approach. **(ii)** *Pinning properties of C-boson*. In the glass state, the equilibrium positions of the clusters of z-$a_0$ and z-$c_0$ clusters in **Figure 5** do not change, and the spatial orientation of the 16 angle lines of $a_0$ remains constant, except for the time when $a_0$ is sequentially projected and overlapped into its four adjacent HSM fields. This property forces C-bosons between $a_0$ and $c_0$ to repeatedly appear in the same spatial position during relaxation time $\tau_i$ of time $t_i$. In the glass state above $T_k$ temperatures, $a_0$ and $c_0$ can be in a "non-changing equilibrium position" in space for a longer relaxation time of $\tau_8$ (at slightly above $T$k, $\tau_8$ tends to infinity), which supports the presence of so-called "pinning particles"[30] in random first-order transition. **(iii)** *Potential energy landscape.* In the one-electron approach, "Classical depictions of the potential energy landscape focus on basins and metabasins which includes multiple basins, with the α-relaxation to connect different metabasins and the *β*-relaxation to connect basins"[31], and the energy difference between different basins is very small, which becomes a difficult problem for potential energy landscapes**.** The two-electron theory states that the potential energy landscape in the glass state refers to the inverse cascade and cascade motion of 2D $V_i$-clusters from small to large. Each molecule in the cluster jumps on a cage of the same scale consisting of 4×16 1/16 equipotential points, so the potential energy of different sized clusters is about the same, but the larger cluster contains more cages and therefore has slightly higher potential energy than the smaller cluster. The one-electron approach can only consider the inverse cascade motion of clusters of different sizes as "fractal"[32]. **(iv)** *Heterogeneity*. density, and kinetic heterogeneity in the glassy state is a hot topic. Zhang et al. used correlated electron microscopy at a sub-nanometer resolution to "visualize spatially heterogeneous dynamics."[33] The observations relied on the theoretical model used to sample the information. As each magic interface is an overlap of two HSMs with the largest amplitude in the system, the distribution of the magic interface is the distribution of the maximum vibrational energy (1/16) in the system, and the number density of C-bosons in clusters of different scales is different. In addition, the number of repetitions of C-bosons at different spatial positions also differed. These heterogeneous phenomena, in turn, support the fact that 2D clusters of different scales in the solid state are inverse-cascade and cascade models of simple hydrodynamic energy fluctuations, as described by the renormalization theory for $n$ = 0. **(V)** *Broad relaxation time spectrum.* The two PCPs in **Figure 5** with 15 0.01 step jumps are the origin of the "terahertz"[34] boson peak, and support Tomoshige N et al.'s view that "a positive correlation between the boson peak, shear elasticity, and the glass transition temperature."[35] In fact, the relaxation time for the production and disappearance of the C-boson can be taken any value, because the two-electron approach only considers when and how the magic interface-boson is continuously excited with temperature completes one to nine closed loops in the inverse cascade. **(vi)** *Excitation energy of C-boson* (CEP interface excited energy): The energy of the C-boson, in addition to the 16 pairs of repulsion electrons, has 16 pairs of electron orbital magnetic moments, is a material parameter that depends on the potential well energy $\varepsilon_0$ of the HSM in the mean-field, which is approximately one-eighth of $\varepsilon_0$, and he typical well energy of the polymer is 51.6 K in the WLF equation. Thus, the energy of the C-boson is approximately 6.5 K (~ 0.56 meV), which is consistent with the observed peak of a low-temperature boson at approximately 6 K[27]. The experimental boson peak in strain glass is approximately 10 K[36], whereas the boson peak range in metallic glass is 5– 20 K[37], which supports the theoretical energy range of the CI bosons.

### 4. *Two-electron trap state.*

The $a_0$-PCP (and synchronized $c_0$-PCP) shown in **Figure 5** must climb out of the two-electron trap state at an angle between $\lambda_\beta$ = 16th and $\lambda_\gamma$ = 1st. ∠Trap ≈ 0.019° ≈ 0.02°). During a 4-diagonal closed-loop jump in the cage, $a_0$-PCP must pass through four trap states.

### 5. *Quasi-resonant and shared resonance modes*.

The PCPs of all HSMs in the soft matrix in the local region follow the PCP of the central HSM $a_0$-along the same 4-diagonal path in their respective 2Δd cubic-lattices (**SM 6, Figure S7**), Note: The cubic lattice has four equivalent 4-diagonal closed-loop paths, **SM 6**). Because the 4-diagonal with a relaxation time of $\tau_i$ contains four trap states, such states in an infinite number of local regions of the system are quasi-resonance modes with a relaxation time of $\tau_i$. The nine relaxation times required for nine closed-loop transitions at different scales can explain the origin of the two-level (ripplon level) in the random first-order transition (RFOT)[26]. Obviously, this newly discovered quasi-local resonance mode with trap states will have a more wonderful evolutionary prospect than all nonlinear vibrational modes. The nine quasi-relaxation modes of the cluster-scale magic-interface boson-trap closed loops in the two-electron theory are the most likely "shared resonance"[38] modes in the "resonance mechanism of brain consciousness"[38].

## E. Collective jumping patterns of molecules in the glass transition

### 1. *The physical significance of the critical temperature $T_c$ in mode-coupling theory*.

There are two modes of collective jumping of molecules during glass transition. The first is the macroscopic percolation transition caused by the soft matrix in different directions between different local regions jumping one after another. The energy of the soft matrix excited by each local region is $k_BT_g^\circ$, which is $k_BT_g$ after averaging all local regions. In the second case, the soft matrix of each local region is at the maximum density in mode coupling theory. Because the density of a single soft matrix in the local region is the same as the density of the $N$ soft matrices that jump sequentially in the local $z$-direction in **Figure 6b**, the energy of the latter is not $k_BT_g^\circ$ but $E_{jump}$. In this case, we should consider the balance between the probability of occurrence $\hat{p}_+$ and the probability of disappearance $\hat{p}_-$ of the "pipe wall" composed of $N$ sequential soft matrices with energy $k_BT_g^\circ$. $\hat{p}_+ = \hat{p}_-$ is the balance between the creation and disappearance of the tube wall in the two-electron approach during glass transition.

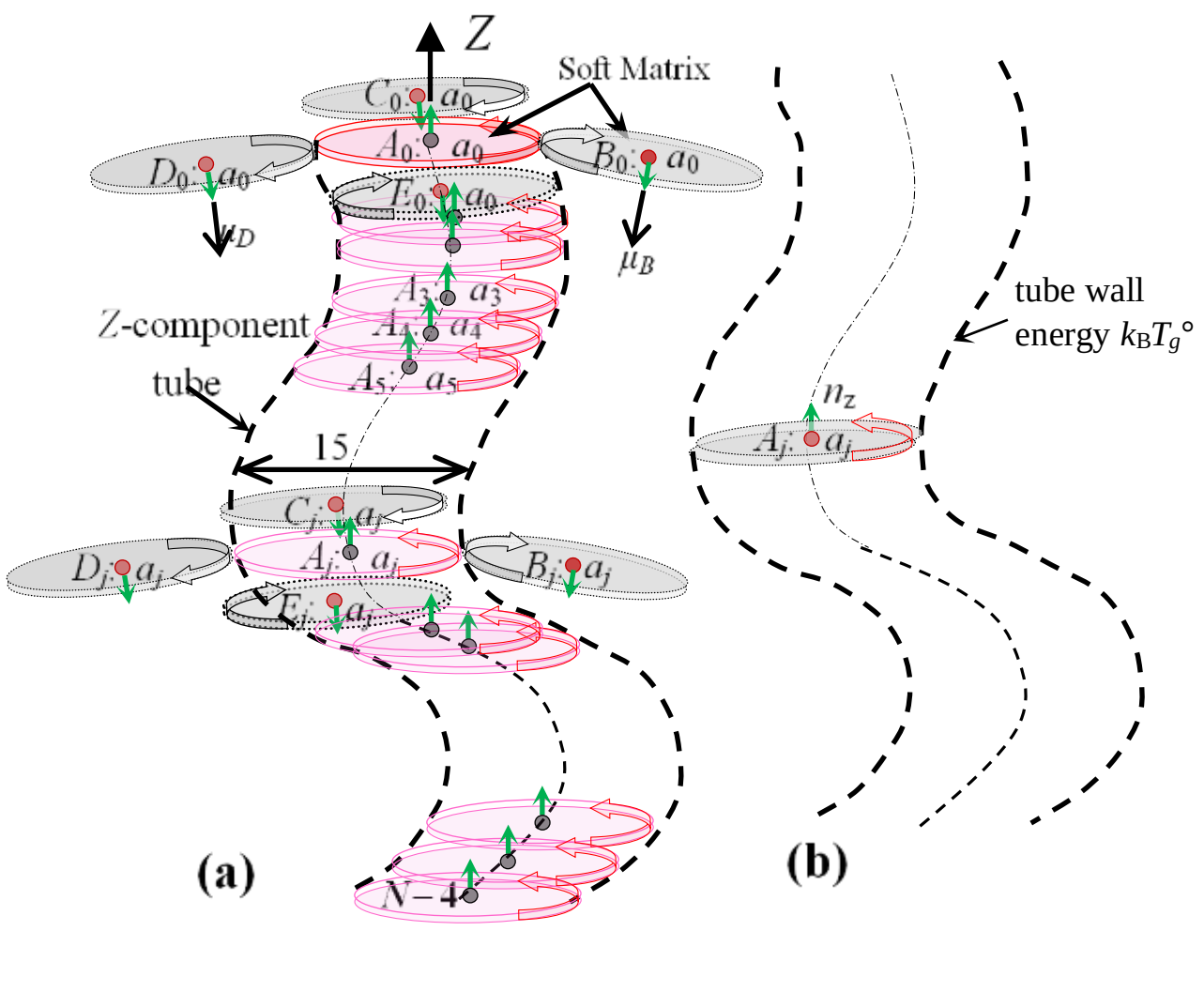

**Figure 6.** Collective motion mode of all molecules in a local region. **a.** All $N$ soft matrices in each local

area that have been excited and will be excited jump $n_z$ steps in the same $z$-direction in sequence. The $z$-direction C-bosons in the $N$ $V_8$-loops form a dynamic $z$-component tube "wall" of length $N$ whose energy, expressed in terms of temperature $T$, is the same as the energy of $a_0$-$V_8$-loop, both are $k_B T_g°$. This correlation is represented by **Figure 6b**. **b.** The probability that the $V_8$ ($a_j$) loop of $a_j$ occupies the tube wall ($N$ associated $V_8$-loops of the $N_z$-chain) is $1/N$. The central vacancy O ($a_0$) of the $a_0$-soft matrix is a +$z$-axial interface soliton formed by all $z$-axial C-bosons, and shared by the central HSMs of $N$ soft matrices on the chain. In **Figure 6b**, the share of $a_j$-HSM in this interface soliton is $n_z$.

Owing to inverse cascading, the energies of the CEP interface excited states in the $a_j$-soft matrix evolve into the $V_8$-loop energy $k_B T_g°$ of the soft matrix. And $V_8$-loop energy $k_B T_g°$ can be equivalent to a "loop-flow" composed of $\alpha_{Lg}$ (= $k_B T_g°/\varepsilon_0$) equivalent particles ($\alpha_1$, $\alpha_2$, $\alpha_3$ ... $\alpha_{Lg}$), **SM 4, Figure S5**, each equivalent particle has a unit DoF energy of $\varepsilon_0$ and is located on its own equivalent chain length $N_z$ (a $z$-component chain with a chain length of $N$). Because the energy of the $V_8$ ($a_j$) loops of the $N$ chain unit $a_j$ on chain $N_z$ is still numerically $k_B T_g$, the $\alpha_{Lg}$ equivalent particle can be equivalent to the association of $N$ $V_8$ loops; therefore, the generation probability $\hat{p}_+$ and disappearance probability $\hat{p}_-$ of the tube in **Figure 6b** are **(SM 4)**

$$\hat{p}_+ = (1/N_z)^{k_B T_g/\varepsilon_0}, \quad (8)$$

$$\hat{p}_- = (n_z)^{k_B T_m/\varepsilon_0} \quad (9)$$

Because each soft matrix shares one DoF energy $\varepsilon_0$ of the central vacancy O ($a_0$) interface soliton of the $a_0$-soft matrix and walks $n_z$-steps along the $z$-axis during its disappearance, the disappearance probability $\hat{p}_-$ of *the tube wall* is also the disappearance probability of $a_j$-HSM occupying the "unit vacancy energy$\varepsilon_0$" $n_z$ share. Here, the disappearance of the $a_j$-soft matrix is caused by the generation of its four neighborhood $B_j$-, $C_j$-, $D_j$-, and $Ej$-local field soft matrices in **Figure 6a**, and the energy involved is $k_B T_m°$. From $\hat{p}_+ = \hat{p}_-$ (this is how the energy fluctuations in the glass transition are balanced), we obtain:

$$n_z = (N_z)^{-T_g/T_m} \leq 0.036 \quad .(10).$$

$n_z$ is the step size and is the number of DoFs for $a_j$ to walk. Thus, the energy that excites the collective jump of molecules in the glass transition can only be the energy of the *magic- interface boson solitary wave*, which causes $a_0$, $a_1$, $a_2$, …, $a_{N-1}$-soft matrices to jump individually in *the same* $n_z$ (≤ 0.036) *steps* and can only be measured by the number of DoFs required in the $z$-direction. For flexible chains with a chain length of 200 critical molecular weights (including small-molecule systems), the DoF number $N^*$ required for $N$soft matrix jumping $n_z$ steps is $N^* \approx 200\, n_z \approx 7.2$. Thus, $E_{jump}$. is approximately 7.2 $\varepsilon_0$, whereas $k_B T_g$ is approximately $20/3\varepsilon_0$ 10 ≈ $6.67\varepsilon_0 < E_{jump}$. Therefore, mode-coupling theory provides a $T_c$ temperature greater than $T_g$.39 The 3D space can be embedded in three 2D spaces corresponding to three 2D soft matrices (**SM 1**, **Figure S1**) of $a_0$-HSM in three independent entangled travel directions in the liquid state. In this case, $a_j$-HSM contains three MIEVs (**SM 5, Figure S6**). This explains the number of DoFs required for the complete free diffusion of a chain with a chain length of $N$ mentioned in the introduction: $N^* = N_z^* \cdot N_x^* \cdot N_y^*$.

### *2. The origin of Johari-Goldstein $\beta_{JG}$ relaxation.*

Another issue related to $T_c$ is the origin of the Johari-Goldstein $\beta_{JG}$ relaxation that occurs around 1.2 $T_g$40, which is also a hot research topic. "Probing its microscopic properties is a crucial step for a complete understanding of glass glass-transition"41. "Unveiling the microscopic mechanism underlying the $\beta_{JG}$ process is not only a crucial step toward a complete theory of glass transition, but is also of significance for many technologies and practical applications"42. As mentioned above, after $T_c$, when all the soft matrices on the $N_z$-component chain has been excited, the movement of the molecule that continues to be excited begins with the excitation of the *x*- (or *y*-) axial soft matrix. Thus, the excitation energy $k_B T_{JG}$ is $E_{jump}$ plus the DoF energy $\varepsilon_0$ in the *x*-axial, so $k_B T_{JG} \approx 8.2\ \varepsilon_0 \approx 1.2\ k_B T_g$. The time required to generate $x$-$V_8$ ($a_0$) soft matrix is the relaxation time for $a_0$-PCP to complete the ninth closed loop on the 4-diagonal around the *x*-axis (**SM 1**, **Figure 1**). Because two of the 4-diagonal closed loop around the *x*-axis have been excited in the 4-diagonal closed loop around the *z*-axis, the time required for the 4-diagonal closed-loop of the $a_0$-PCP around the *x*-axis will suddenly decrease, corresponding to $\beta_{JG}$ relaxation.

## F. Physical image of the glass state

### *1. Equilibrium mode of kinetic energy and potential in glass state, Kauzmann paradox*

"The existence of an ideal glass and the resolution to the Kauzmann paradox is a long-standing open question in materials sciene"[43]. Glass states in two- and three-dimensional spaces have also been a hot topic in the exploration of glass formation. "Glass transitions may be similar in two and three dimensions"[44]. Two-electron glass state theory answers these two questions in a straightforward manner. Because of the closed-loop requirement of $n$ = 0, the 4×16 eigenvalues that appear sequentially in the 2D $V_0$-cluster are actually a transformation from the 3D positional disordered jumps of the five $\sigma$ hard spheres in $\sigma_0$ to the ordered jump of positive and negative charges at 4×16 pairs of coupling discrete points in the 2D overlapping space.. In other words, the $n$ = 0 closed-loop operation is the equilibrium between the directional potential energy of the C-bosons in the 2D $V_0$-cluster and the disordered thermal vibrational kinetic energy of the $\sigma$ hard sphere on the 3D $\sigma_0$ scale. Starting from the absolute temperature, at the point where the vibrational energy of the 5-molecule of the local thermal fluctuation reaches the maximum value (1/16), the eigenvalues of the two overlapping hard spheres appear in turn, and 2D $V_0$–clusters in different directions are formed in the form of dynamic cubic lattices at different positions in 3D space. In 3D space, these $V_0$-clusters are an inverse cascade from small to large along the *z*-direction of the first 5-hard sphere cluster in the local area; thus, the potential energy of the *z*-axial eigenvalue in the increased 2D cluster is always in equilibrium with the disordered kinetic energy $k_B T$ in 3D space. Therefore, the Kautzmann paradox does not exist for the two-electron approach.

### *2. Kauzmann $T_K$ temperature and glass viscosity tend to infinity*.

"A central puzzle in glass physics is why a glass-forming liquid becomes so viscous before forming a glass"[45], In the two-electron theory, the vitreous viscosity problem becomes simple; at any temperature from solid to liquid, it is a soft matrix jump rather than a molecule. Therefore, the number of soft matrices embedded in disordered molecular systems in the glass state remains unknown. The difference in the state at different temperatures ($T_K$, $T_g$, $T_m$) is caused by the difference in the number density of the soft matrix in the local region. When the number density is five, a melting transition occurs; when the number density is one, a glass transition occurs; and when the number density tends to zero, it is the Kauzmann $T_K$ temperature, and the viscosity tends to infinity.

### *3. The origin of solid-state rigidity*

The two-electron theory also solves an old problem in solid-state physics that has not been solved for nearly a hundred years, that is, the origin of solid-state rigidity. "Amorphous solids, or glasses, are apparently rigid as a crystalline state of matter, but at the same time, disordered as a liquid state. Such a combination of rigidity and disorder remains a fundamental open question in condensed matter physics and materials science despite serious efforts over the years"[46]. Tong et al. proposed a new perspective of the "self-organization" of the system from the perspective of mechanics, and the two-electron theory classifies this "self-organization" as the inverse cascade of clusters. When the temperature is below $T_K$, the system is filled with $V_0$ to $V_7$-clusters of repulsive electron pairs, which is the origin of the rigidity of the solid state, and any tiny displacement of the solid state must be an inverse cascade to form a large number of soft matrices and reach the $k_BT_m$ energy per unit volume and consume the $k_BT_g$ energy per unit volume.

***4*. Various structures of glass states**.

High-temperature superconductivity is the transition from a glassy state to a superconducting state; therefore, two-electron theory can provide superconductivity scholars with a new perspective on the transition from the glassy state to the superconducting state. In particular, the fact that all magic-surface bosons in the 2D $V_7$ clusters are oriented along the *z*-axis, which is the simplest mechanism for the strong correlation of the 2D electrons**.** Similar to ferromagnets, there are different paths (magnetic hysteresis loops) from low to high temperatures and from high to low temperatures: inverse-cascade from glass state-1 at low temperature $T_1$ to glass state-2 at $T_2$ (above $T_k$ and below $T_g$), and then continue inverse-cascade to temperature $T_3$ above $T_g$; then, cascade from $T_3$ to $T_2$, where the glass state is recorded as glass state-3, and then cascade to temperature $T_1$ to give glass state-4. There are two cases to be aware of in this inverse-cascade -cascade "hysteresis loop," In the first case, at the same temperature $T_2$**,** the structures of glass state-2 and glass state-3 are different, glass state-3 is called "supercooled liquid state," which contains more soft matrices than glass state-2, and when it relaxes to glassy state-2, it can be speculated that it will follow the law of abnormal exponential relaxation of the glassy state[10]. This suggests that high-temperature superconductivity does not occur in the temperature range above $T_k$ In the second case, if the $V_i$-cluster has two energies, at the same temperature $T_1$, the $V_i$ cluster in glass state 1 chooses a lower energy, whereas the $V$-cluster in glass state-4 chooses a higher energy, which is also one of the necessary conditions for high-temperature superconductivity to eliminate the two-electron trap state. Section **IV.D.*2*.** In addition, the cooling method from temperature $T_3$ to temperature $T_1$ also affects the structure of glass state-4. Under rapid cooling conditions, all the molecules in the localized region choose the same 4-diagonal closed-loop path on their cages to respond quickly to external temperature changes. Under very slow cooling conditions, each molecule in the local region can freely choose a 4-diagonal closed-loop path along the *x*-, *y*-, and *z*-axes of the cage. There are four path choices in each direction, resulting in $V_7$-clusters in different directions in the local area intertwined with each other. The cascade condition is no longer satisfied (which is *the microscopic mechanism of glass annealing*), and the glass state at low temperature $T_1$ has $V_7$-clusters intertwined in all directions, which is the microscopic mechanism that *eliminates the brittleness of the glass state*. Therefore, quenching must be performed in order to prepare high-temperature superconducting materials.

***5 Clustered bond and aging of materials***.

The aging mechanism of materials has also been a topic of concern, Till Böhmer et al. determined that physical ageing is the "irreversible processes in glassy materials resulting from molecular rearrangements"[47]. This discussion of the nature of material aging is only half of the story and is not sufficiently thorough. To understand the aging mechanism, it is necessary to first understand the *clustered bond* between the molecules in the mean field. According to the two-electron theory, the interface of the two molecules in the mean field is a magic interface, the interaction between the two molecules is a C-boson, and the clustered bond energy of the two molecules is $k_BT_g$ (the energy to separate the two molecules when the cage of the molecules of the material is a 2Δd cubic-lattice). Therefore, the two-electron theory gives the microscopic mechanism of material aging without any suspense: In the mean-field, when a large number of neighboring two molecules have a spacing greater than $d_L$*, the two HSMs are unable to project-overlap to produce the magic interface, and the loss of a large number of cluster bonds between molecules is the molecular mechanism of the material's ageing,

## G. Closed loop of four magic interfaces in the form of an equilateral hexahedron

### *1. Dynamic equilateral hexahedron embedded in glass / liquid*.

When further exploring why the soft matrices of different materials have different Hamiltonians, $H = k_B T_g$, it is found that the equilateral hexahedron is a more general geometric shape that satisfies $n = 0$, as shown in **Figure 7**. Because there are two independent variables in the three angles $\Omega_x$, $\Omega_y$ and $\Omega_z$ that can take any value, there are countless types of equilateral hexahedrons satisfying $n = 0$ in the glass/liquid states provided that the side length of the equilateral hexahedron is $1 + d_L$* between two adjacent HSMs in the cluster. In the case of a liquid/glass equilateral hexahedron, **Figures 2–7** still holds. However, the disordered vibrational energy $k_B T_g$ required to balance the soft matrix-ordered energy $k_B T_g°$ of the equilateral hexahedron in the *z*-direction is:

$$k_B T_g \cos\Omega_z = k_B T_g^{\circ} \tag{11}$$

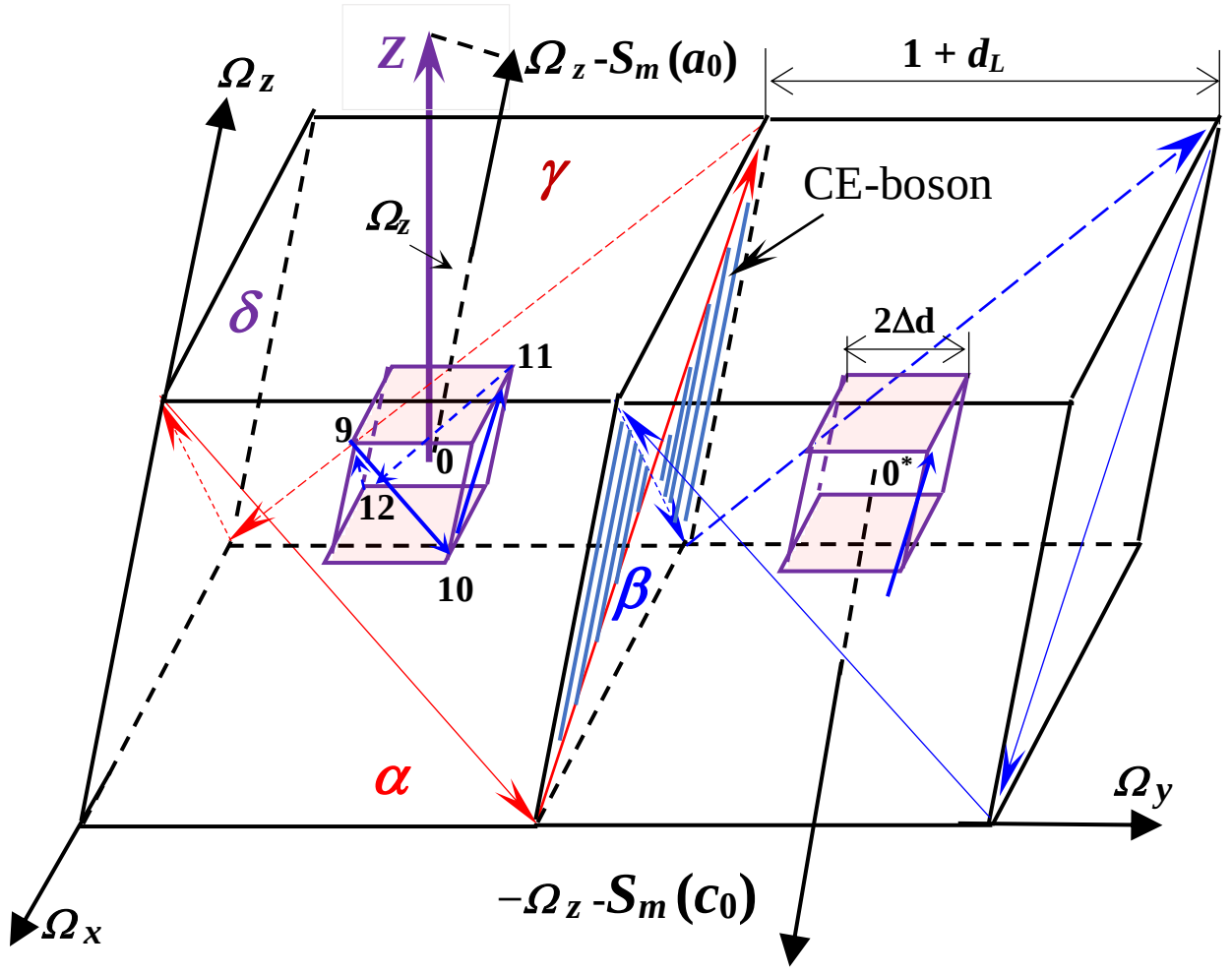

.

**Figure 7.** Magic interface equilateral hexahedral lattice vector illustration**.** The two diagonals of each face (square or rhombus) of an equilateral hexahedron are orthogonal, satisfying $n = 0$ conditions of adjacent two-HSMs SASC magic-interface and their PCP pair and electron pair synchro-orthogonal-coupled-jumps.

This means that while the molecules form HSMs in the L– J potential in the mean field, they also spontaneously establish four L– J potentials in the $\pm\Omega_x$- and $\pm\Omega_y$-directions according to their chemical molecular structure, such that each HSM has four 1/16 equipotential planes in the form of a 2Δd dynamic cubic-lattice /equilateral hexahedron in the mean field. Note that in **Figure 7,** if all four equipotential faces in each cage in the local area have the same torsion angle around the *z*-axis, the condition $n = 0$ is still satisfied, which is *a necessary condition for high-temperature superconductivity*.

### *2. 2D dynamic magic interface lattice vector diagram of the polyethylene chain unit*

Based on the trap-state properties**,** the dynamic 2D molecular magic-interface lattice of the polyethylene (PE) chain unit (**Figure 8a**) and its HSM model (**Figure. 8b**) are shown in **Figure 8c** and **Figure 9.**

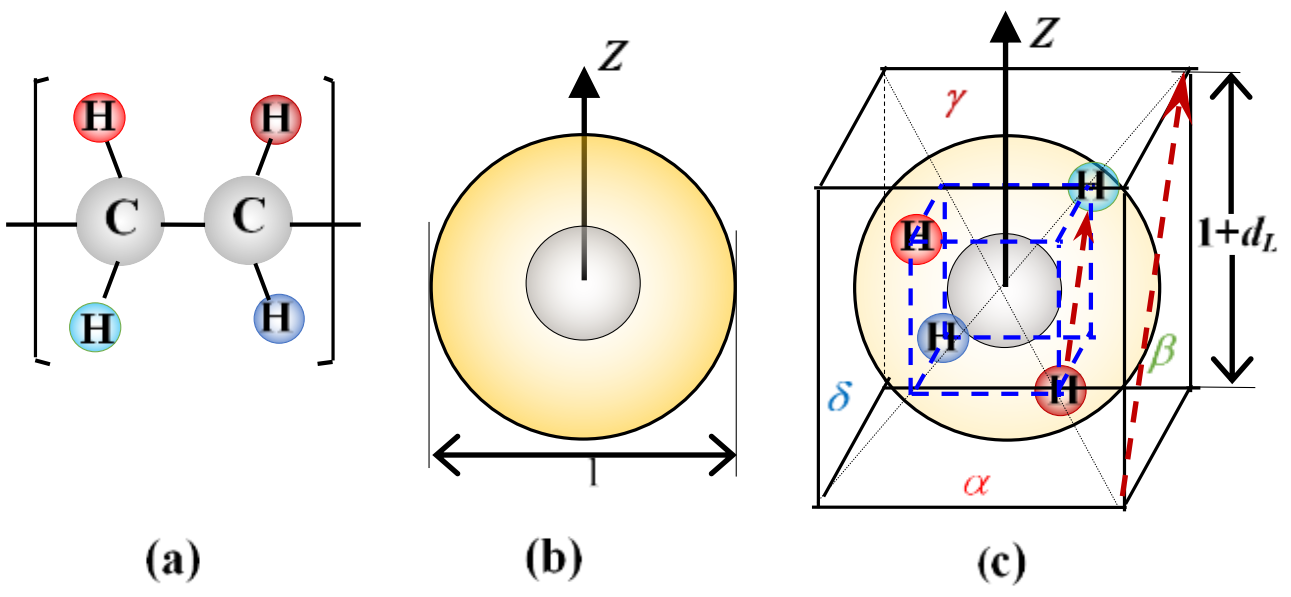

**Figure 8**. PE chain unit C2H4 HSM model and dynamic cubic-lattice in clustering**.** a. Chain unit structure with four hydrogen atoms at different chemical positions marked with different colors. **b**. HSM model in the mean-field. **c.** The four hydrogen atoms in clustering appear sequentially at the four starting positions of the larger cube lattice (concentric and synchronous with the 2Δd cube lattice).

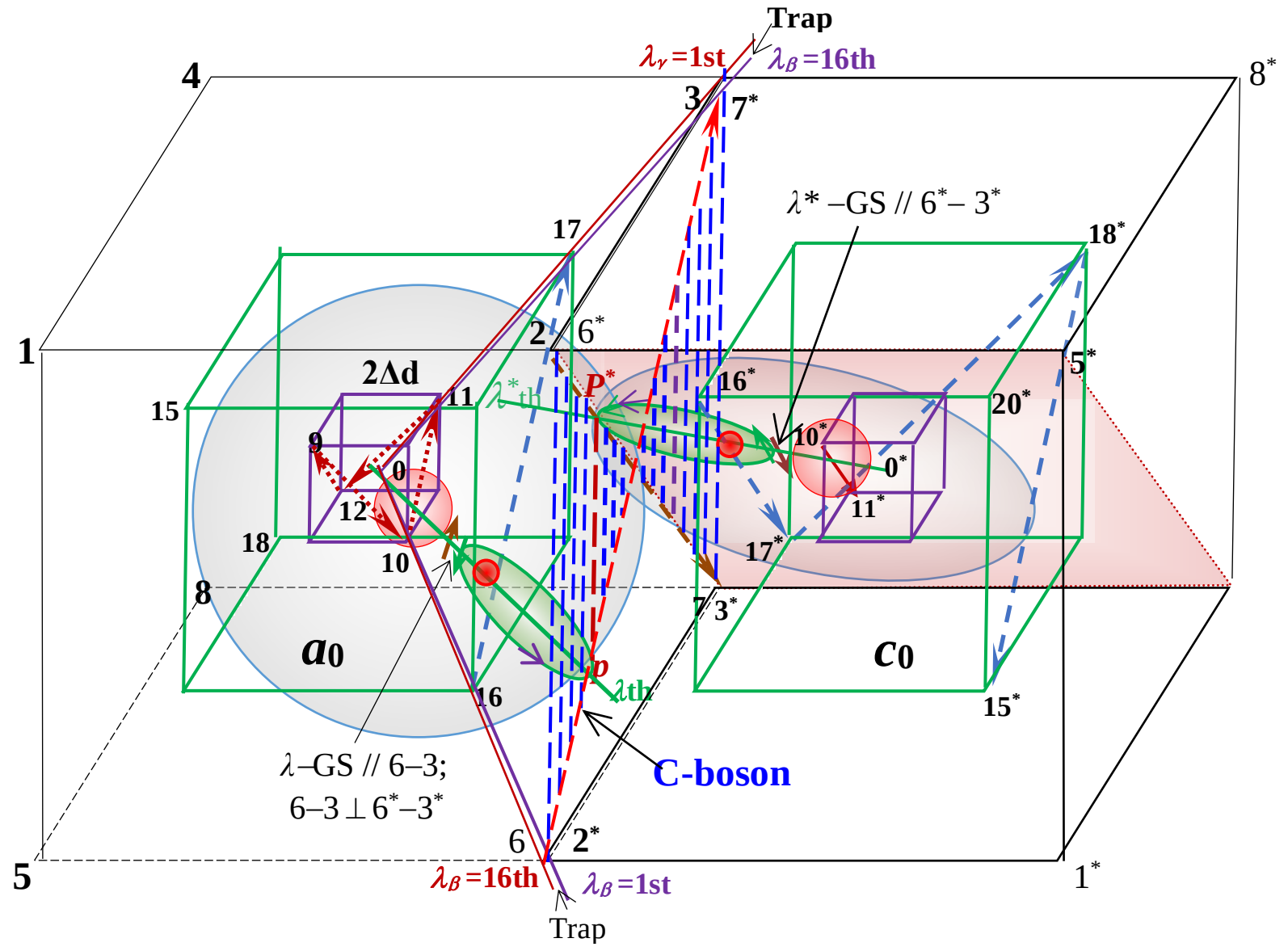

**Figure 9**. Clustering diagram between adjacent chain units of polyethylene in the glass state**.** Due to the symmetry broken of HSM in the overlap, the two positive charges in the C–H bond, two red balls, one big and one small, are located at this $\lambda$th angle line, and two ground state (GS) electrons are defined as states parallel to two orthogonal diagonals.

### *3 2D dynamic magic interface lattice vector diagram of the polypropylene chain unit*

The polypropylene (PP) chain unit is -[-CH2--CH (CH3)-] ( **Figure 10a**. The HSM is shown in **Figure 10b**, and its dynamic equilateral hexahedron is shown in **Figure 10c**. There may be more than four H-PCPs (positively charged particles of hydrogen atoms or groups) in a chain unit, and the rule is to select the H-PCP that is most likely to escape an electron in a chemical reaction. The $T_m$ and $T_g$ of polypropylene are denoted as $T_m$ (PP) and $T_g$ (PP), respectively, because the difference between $T_m$ (PP) and $T_g$ (PP) is close to the difference between $T_m$ (PE) and $T_g$ (PE) of polyethylene; both $\Omega_\beta$ and $\Omega_\gamma$ in the equilateral hexahedron of polypropylene (in **Figure 6**) can be approximately 90°. In addition, attribute the higher $T_g$ (PP) of polypropylene than $T_g$ (PE) to $\Omega_z$ being less than 90°. Taking a conventional linear amorphous polyethylene $T_g$ (PE) value – of 78 C° and polypropylene $T_g$ (PP) value – of 10C°, we roughly obtained $\cos\Omega_z \approx T_g$ (PE) / $T_g$ (PP) ≈ 195 K / 263 K, $\Omega_z \approx 42°$**.** Here, the equilibrated hexahedral excluded volume of polypropylene is approximately equal to its HSM volume, whereas the excluded volume of polyethylene is 1,35 times the volume of polyethylene HSM, which is the so-called steric hindrance effect of C-H3 in polypropylene.

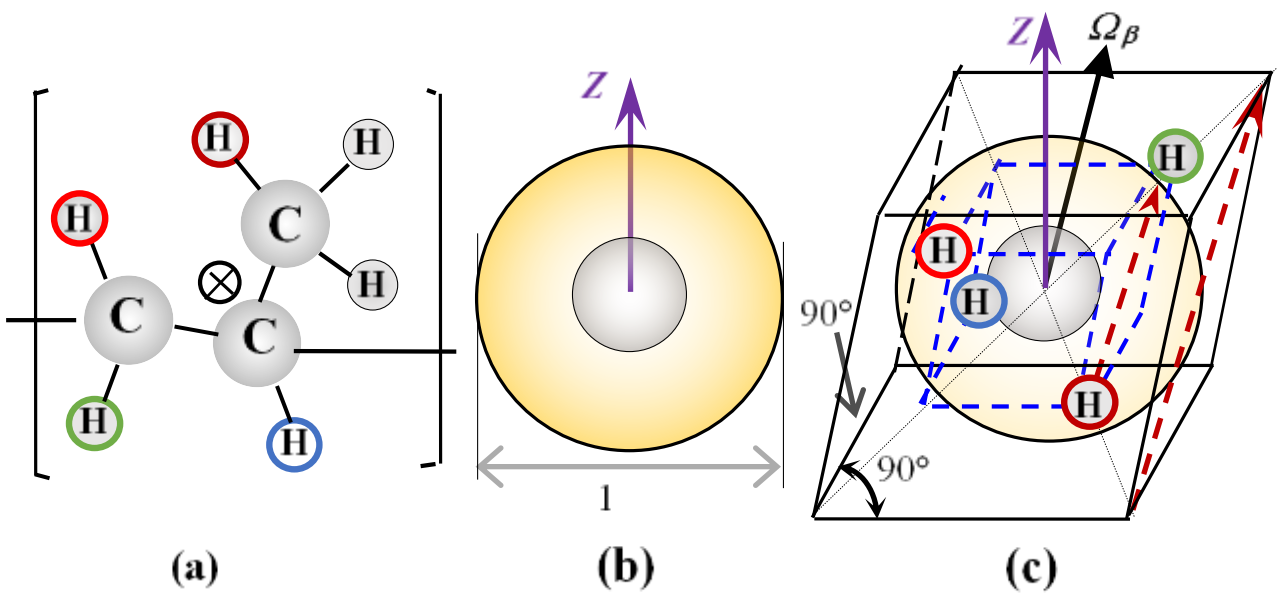

**Figure 10**. Amorphous polypropylene chain element HSM model and its dynamic equilateral hexahedron. **a**. The molecular structure of the chain unit. Four of the six H-PCPs (different colors) are selected to participate in clustering. The ⊕ symbol denotes the *z*-axis position through the center of mass of the chain unit. **b.** HSM model**. c.** The four H-PCPs in the cluster appear sequentially at the four starting positions of the larger equilateral hexahedral lattice (concentric with the 2Δd equilateral hexahedron).

The chain unit -[-CH2--CH (CH3)-]- (**Figure 10a**) and its HSM (**Figure. 10b**), are shown in **Figure 10c**. There may be more than four H-PCPs (positively charged particles of hydrogen atoms or groups) in a chain unit, and the rule is to select the H-PCP that is most likely to escape an electron in a chemical reaction. Because the difference between $T_m$ (PP) and $T_g$ (PP) of polypropylene (PP) is close to the difference between $T_m$ (PE) and $T_g$ (PE) of polyethylene, both $\Omega_\beta$ and $\Omega_\gamma$ in the equilateral hexahedron of polypropylene (in **Figure 6**) can be approximately 90°. In addition, attribute the higher $T_g$ (PP) of polypropylene than $T_g$ (PE) to $\Omega_z$ being less than 90°. Taking a conventional linear amorphous polyethylene $T_g$ (PE) value – of 78 C° and polypropylene $T_g$ (PP) value – of 10C°, we roughly obtained $\cos\Omega_z \approx T_g$ (PE) / $T_g$ (PP) ≈ 195 K / 263 K, $\Omega_z \approx 42°$**.** Here, the equilibrated hexahedral excluded volume of polypropylene is approximately equal to its HSM volume, whereas the excluded volume of polyethylene is 1,35 times the volume of polyethylene HSM, which is the so-called steric hindrance effect of C-H3 in polypropylene.

#### *4. 2D dynamic magic interface lattice vector diagram of arsenic trisulfide (As2S3).*

Small-molecule arsenic trisulfide (As2S3) is one of the most commonly used materials for studying glass transition.. Using the $T_k$, $T_g$ and $T_m$ data for As2S3 in the literature, the 2D dynamically ordered structure of the equilateral hexahedron of As2S3 in the mean fields qualitatively displayed**.** For As2S3, take $T_k$ = "97C°"48 = 370 K, $T_g$ = "175C°"49 = 448 K and $T_m$ = "583 K"50 since $(T_m - T_g)$ K / 4 ≈ 34 K, if we take $\varepsilon_0 = 34\ k_BT$, then $T_g° \approx$ 370K + 34K = 404K. It is determined that in the equilateral hexahedron of As2S3, $\Omega_\beta$ and $\Omega_\gamma$ are 90, ° and $\Omega_z$ is less than 90°. From Equation (9), we obtain $\Omega_z = \cos^{-1}(T_g°/ T_g) \approx 25.6°$.

### *5. Two ordered fluctuation energies in a disordered system*.

The new concept of C-bosons enables the emergence of two ordered energies in each disordered system: $k_BT_g°$, the energy that generates a $\Omega_z$ -axial soft matrix; and $k_BT_m°$, the energy that disappears from the $\Omega_z$ -axial soft matrix. For flexible polymer chains, $k_BT_g° = k_BT_g$, $k_BT_m° = k_BT_m$. In general, $k_BT_g° \neq k_BT_g$, $k_BT_m° \neq k_BT_m$. There is a factor cos $\Omega_z$ between $k_BT_g°$ and $k_BT_g$. However, the following formula holds

$$T_g° / T_m° = T_g / T_m \qquad (12)$$

This is because $T_g° / T_m°$ eliminates this factor. This result holds in Equation (10), and $n_z = (N_z)^{-Tg/Tm} \leq 0.036$ for all molecular systems.

# IV. APPLUCATION OF THE TWO-ELECTRON THEORY

## A. Magic Interface-boson soliton replaces free volume

The new concept of the magic-interface boson soliton fundamentally updates the definition of "free volume" in the early days of the glass transition. This is also the idea conceived by de Gennes in his simple picture of structural glass4**:** the need to replace free volume with a suitable new concept**.** In this study, the vacant "free volume" in the center of the *z*-axis soft matrix is a soliton, and **Figure 6** depicts the mode of motion of the collective molecules in the glass transition as a solitary wave in the form of a soft matrix that consists of all directional magic interface- bosons that appear sequentially in the local region. The vacancy volume is a dynamic cubic lattice/equilateral hexahedron with an edge length of 1+ $d_L$*. It should be pointed out that the reason why the "vacant volume" contained in the *z*-axis soliton can be shared by the *z*-axis $a_j$ molecule in **Figure 6** in the inverse-cascade-cascade mode of the glass transition. That is, the "empty volume" solitons in the form of a cubic lattice/equilateral hexahedron consist of four magic interface-bosons around $a_0$-HSM nine times, while the four magic interface-bosons of the subsequent $a_1$-HSM have been around $a_1$ eight times, and so on, and all the bosons are strongly correlated in the *z*-axis, thus obtaining the solitary wave mode This property also applies to the exclusion volume. The HSM in the mean field is not the excluded volume of a molecule. The new concept of C-boson allows for a strict definition of the excluded volume of a molecule. The interaction between two adjacent molecules can occur only in a C-boson, and the excluded volume is a dynamic equilateral hexahedral composed of four magic interface-bosons that emerge sequentially.

## B. The glass transition is a confluence of kinetics and thermodynamics

"The confluence of both the thermodynamic and the kinetic dimensions of the liquid ↔ glass transition has presented one of the most formidable problems in condensed matter physics"3. In one-electron approaches, there has been controversy as to whether glass transition is kinetic or thermodynamic, and recent studies have shown "a fundamental correlation between micro-dynamics and thermodynamics"51. The magic interface-boson provides a simple statement. The fact that the magic interface bosons appear sequentially to form a soft matrix suggests that glass transition is a combination of kinetics and thermodynamics. This study clarifies that $n = 0$ is a universal theory that describes quenching disorder from disorder to order. Glassy quenching disorder of the magic interfaces began near the absolute temperature and continued to emerge as the temperature increased. Glass transition occurred when 320 magic interfaces appeared. Because the confluence of kinetics and thermodynamics is one of the most fundamental concepts in life sciences, it is predicted that a series of new concepts in glassy two-electron theory will also play a role in the evolution of physics to biochemistry.

## C. Theoretical proof and simplification of the redefined Born-Oppenheimer approximation in quantum chemical reactions

The discovery of the two-electron theory provided an opportunity to prove the Bonn-

Oppenheimer (B-O) approximation hypothesis. "The Born–Oppenheimer (BO) approximation is the bedrock of quantum mechanical calculations of atomic and molecular systems. However, these systems require departure from the BO approximations"52. The BO approximation hypothesis is widely used in quantum chemistry and chemical-reaction kinetics. This hypothesis, which is almost 100 years old, refers to the position where the nucleus can see its electrons reach equilibrium when displaced. This approximation makes it possible to separate the motion of the nuclei and the motion of the electrons, and forms the basis for most electronic structure methods. In recent years, atomic-electron detection techniques with higher accuracy have revealed that *experimental* results deviate from the original BO approximation, and the theory of "non-BO approximation"53 has been developed. However, the non-BO approximation ignores the fact that most chemical reactions occur in the mean field and there is a two-electron theory in the mean field. In this study, the two-electron theory can also be considered as an exploration of how the electrons of the two molecules synchronously move to the interface during a chemical reaction. As shown in **Figures 5 and 9**, the PCPs in a chemical reaction do not jump randomly, and only when two PCPs synchronously acquire the maximum vibrational amplitude energy (1/16) and jump synchronously along two orthogonal diagonals, can each nucleus see the position of one of its electrons involved in the chemical reaction at the magic interface. In other words, the movement of a nucleus and one of its electrons in a chemical reaction cannot be separated and both are in synchronous motion. a possible solution to the unified BO approximation and non-BO approximation is to slightly modify the original BO approximation: "In a chemical reaction, when the nucleus moves, it can see an emerging new orbital, the orbital electron involving in the chemical reaction is located at the magical interface.” The two-electron theory not only proves the modified BO hypothesis but also greatly simplifies the theoretical calculations of quantum chemistry, in which the two-electron theory allows only 16 pairs of electron orbitals emerging in the two-electron approach to perform bimolecular chemical reactions. Two-electron theory states that the synchronous jump of two PCPs in a chemical reaction is always orthogonal, suggesting that the two-electron theory of $n$ = 0 may also be a mathematical tool for "bioorthogonal chemistry"58. In turn, we can look for a new geometric behavior of $n$ = 0 in bioorthogonal chemistry (the third solution of $n$ = 0) from a large number of bioorthogonal chemistry examples that have been proven to be correct.

## D. Theory of high-temperature superconductivity

### *1. Electron pairing*

Both the high-temperature superconductivity and glass states originate from the new concept of a magical interface boson consisting of two adjacent atoms or molecules in the mean field. Two SASC PCPs of two adjacent HSMs jump in 15 steps along the two orthogonal diagonals, and the 16 $\Omega_z$ azimuth repulsion electron pairs of the CEP of the two HSMs appear in the overlapping magic interfaces. This is electron pairing in high-temperature superconductivity.

### *2. Meissner effect*

The central HSM-PCP jumps along the 4×16 steps of its cage to form two synchronous and concentric $\Omega_z$-axial dynamic equilateral hexahedrons with side lengths of 1.1046 and 0.1103 (a dimensionless unit well energy measure), respectively, which are the smallest $\Omega_z$-axial 2D $V_0$-cluster. Of the nine transitions from the 2D $V_i$- to $V_{i+1}$ clusters, seven glass (including metallic glass)-superconducting transitions may occur. The first hard-sphere $\sigma$ - $V_0$–superconducting transition is classical BCS superconductivity, corresponding to the universal constant $d_L$* for disordered systems, where the one-electron and two-electron theories converge. At this point, the edge length of the lattice in low-temperature one-electron theory, measured in terms of the dimensionless potential well energy, is exactly 1+ $d_L$* = 1.1046. Two synchronously coupled electrons can be spontaneously located at the two vertices of the lattice, which allows the 16th pair of repulsive electrons at the $V_0$ cluster interface to perform 16th step parallel jump transport. High-temperature superconductivity is a two-electron trap-superconducting transition. If each cage in the local region has additional torsional energy in the $\Omega_z$-axis direction, then when cascading from $V_7$ clusters to $V_i$ clusters, the four 16th orbital magnetic moments in the four trap states of the $\Omega_z$-axis of all the cages in $V_i$ clusters can jump out of the four traps by sequentially acquiring torsional energy, realizing the transition from glassy to high-temperature superconducting states. $V_7$-clusters are in the percolation state at $T_k$ temperature, and all $V_i$-clusters remain in the percolation state when $V_7$ clusters are cascaded to $V_i$-clusters (Note that $V_i$-clusters in the inverse cascade are not in the percolation state because the number of $V_i$-clusters at this time is not sufficient to form a percolation). Subsequently, the superconductivity of the $V_i$ clusters appeared in an equilibrium system in which the 4-diagonal closed-loop jumping of each HSM-PCP along the cage was balanced by torsional vibrations of the cage. In the closed-loop jumping of the $a_0$-PCP along the 4×16 equipotential on the cage, the magnetic moments of the 4×17 electron orbits form a complete closed loop around the $z$-axis, which is the microscopic origin of superconducting Meissner diamagnetism (Meissner effect).

### *3. Magic angle*

The reason why the metallic glass state cannot be a superconducting state is that there is a two-electron trap state with an angle ∠Trap = 0.019°. In **Figures 2** and **5**, in the 15 jump of $a_0$-PCP, the magnetic moments of the 16 electron orbitals totaled (90° – 0.019°) around the $z$–($\Omega_z$) axis. $a_0$-PCP jumps from point $q_{\beta\text{-}16}$ to point 11, which is actually $a_0$ jumping from 1/16 repulsion equipotential point $q_L$ of the +$y$ axis L– J potential to 1/16 attracting equipotential fixed point $q_R$ of the-$x$-axis L–J potential. Depending on the structure of the glassy cage (**SM 6**, **Figure S7**), the condition $n$ = 0 is still satisfied when each cage that appears sequentially in the local region has the same additional torsion angle ∠Tors on the $z$-axis, and each of the four interfaces of the cage in turn contains additional torsional energy in the $z$-direction, and the 4-diagonal 4 trapped states can be eliminated sequentially to achieve a superconducting transition.

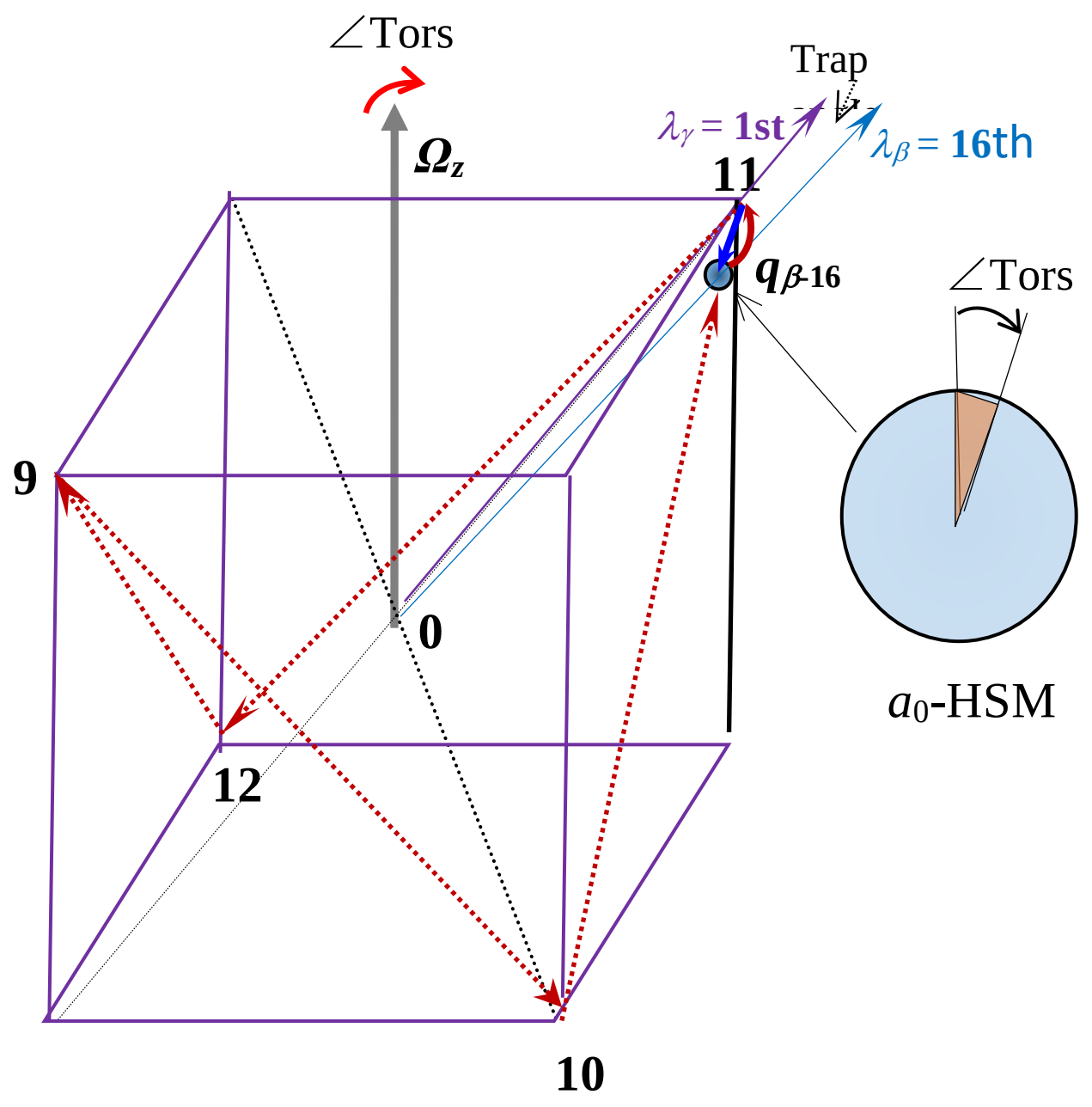

**Figure 11.** Enlarged schematic diagram of $a_0$ cage torsional vibration versus $a_0$-PCP 4-diagonal closed-loop jump balance. The $\Omega_z$-axis of the $a_0$ cage has a small twist angle ∠Tors (red arrow). The length between points $q_{\beta\text{-}16}$ and 11 is the radius of the arc, so this arc has unit torsion well energy. $a_0$-HSM gains

the energy to jump out of the potential well, and, at the same time, the torsion angle of a0-HSM disappears.

The details are as follows: In $\beta$–space between $a_0$ and $c_0$, the two electron orbital magnetic moments of the coupled $\delta_\lambda$–$\delta_{\lambda^*}$ angle lines are still located on the two orthogonal diagonal planes. After $a_0$-PCP makes 15 jumps along a straight line from point 10 to point $q_{\beta\text{-}16}$, the 16 successive electron orbital magnetic moments compress the twisting energy on the surface of 10-11 into the trap angle of $a_0$ and the compression process forces the $a_0$-hard-sphere to rotate an angle ∠Tors (yellow) in the opposite direction along the $z$-axis. At the same time, $a_0$ is no longer a straight line from point $q_{\beta\text{-}16}$, to point 11 but is an arc (red), **Figure 11**. The twisted energy of this arc is equal to the energy of the reverse rotation angle of $a_0$-hard-sphere**.** In **Figure 4**, the unit of measurement for the jump angle $\Delta\Omega \equiv (q_R - q_L)/\sigma$ is "dimensionless radians with an angle of 60°.

The trap angle ∠Trap is not yet an angle around the $z$-axis, unless the arc from point $q_{\beta\text{-}16}$ (the repulsive state $q_L$ corresponding to the maximum torsional energy of $a_0$-hard-sphere) to point 11 (corresponding to the attractive state $q_R$ with zero torsional energy of $a_0$-hard-sphere) is a dimensionless unit radian of the torsion angle ∠Tors measured by the angle ∠Trap, containing a 60° angle, That is, ∠Tors/∠Trap = 1 radian = 60, resulting in ∠Tors ≈ 1.14°~1.17°. This is a universal constant called the magic angle, and its theoretical value compares to Cao Yuan's experimental value: "the torsion angle is about 1.1°"54, which is consistent within the error range. The Cao Yuan's "1.7 Kelvin critical temperature"54 is interpreted in the two-electron theory as the $z$-axial clustered bond energy of the two-electron dynamic equilateral hexahedron hidden between the layers of 2D graphene is only 1.7 $k$T. The unit radian here is the ratio of the unit twist energy $\varepsilon_0$ to the unit torsion radius (the distance between point $q_{\beta\text{-}16}$ and point 11). Thus, the magnetic moment of the 16th electron orbit of $a_0$ obtains the equivalent potential well energy $\varepsilon_0$ with an additional torsion angle, jumps out of the trap, rotates 0.019° around the $z$-axis, jumps from point $q_{\beta\text{-}16}$ to point 11, and the torsional angle of $a_0$-hard-sphere disappears. Therefore, in the closed-loop jump of $a_0$ along the 4-diagonals, the four trap states in the closed-loop path disappear one after another. At the same time, $a_0$- $c_0$ generates a pair of orbital magnetic moments at points 11-11$^*$ in $\beta$-space, and the 16th pair of repulsive electrons will jump to points 11-11$^*$ and parallel to the $z$-axis under an external electric field, forming superconductivity without resistance; at this point, the $a_0$-HSM has a $z$-axial interface-excited spin in the true sense of the word in the form of a cubic-lattice/equilateral hexahedron (the two-electron theory provides the origin of the spin wave of molecules, which is the most difficult to understand in the one-electron theory). Therefore, the goal of the search for a high-temperature superconducting material is to find such a room-temperature superconducting material whose cage has a slow torsional vibrational frequency along the $\Omega_z$-axis at room temperature (relaxation time $\tau_6$, corresponding to the time required for each positively charged particle in the $V_6$-cluster to jump along the 4-diagonal closed-loop of the cage).

# V. DISCUSSION

This paper shows that the disorder of the glass state is simply an irregularity in the average position, and the main focus of the discussion in this paper is that the long-range order of the glass disorder system is a "hidden structure", hidden in **Figures 3** and **4**, as well as in equations (5) and recursions (6-2). Many scholars believe that "liquids and glasses have neither long-range translational nor orientation orders"55, however, the "long-range order" of glass disorder systems is a special way to move from disorder to order in the critical state of the mean-field HSM model. Each HSM in the mean field expresses *the isotropic long-range interaction* of the molecule with a myriad of molecules in the system, and a 0.27% overlap of the two adjacent HSMs along the z-axis causes *the de- isotropic long-range interaction* between the two HSMs and all the HSMs in the local region *along the 16 coupled* $\lambda$–$\lambda^*$ *cluster contact angle lines* (which is why the 16 pairs of SASC $\lambda$–$\lambda^*$ angular lines are called eigenvectors of disordered systems) and generates 16 ordered repulsive electron pairs at the *z*-axial magic interface of the two HSMs. Thus, the molecules in the mean field have 2D $V_0$-cluster vectors, and the inverse cascade of all $V_0$ clusters along the *z*-axis is a long-range ordered interaction, unique to disordered glass molecules. This inverse cascade order in the mean field may be the theoretical source of physicochemical self-organization. Two-electron theory is bound to have a profound impact on the development of all molecular disciplines. Water, for example, in addition to the various crystalline structures determined by quantum mechanics, has a dynamic equilateral hexahedral form consisting of the quenching disordered eigenvalue (**SM 5**) determined by quantum mechanics combined with $n$ = 0 theory, which may play a decisive role in physicochemical reactions. These findings provide new insights for future water research.

In physics, any composite particle consisting of an even number of fermions is a boson, because bosons have integer spins, whereas fermions have odd semi-integer spins. Tichy et al. proposed that " Composite particles made of two fermions can be treated as ideal elementary bosons as long as the constituent fermions are sufficiently entangled;"56 This sufficiently entangled structure had three implications for the glass state. **(i)** Two electrons are coupled electron pairs of two adjacent molecules, each of which is an HSM that interacts with all molecules in the local region; that is, two electrons in two HSMs already contain entangled interactions with the electrons of all molecules in the local region. **(ii)** The two SASC HSMs overlap to form a *z*-axis magic interface, and the two HSMs de-interact with all the molecules in the local region along the *z*-axial 16 pairs of SASC cluster-contact angle lines. The PCPs of the two HSMs jumped along the two orthogonal diagonals, thereby eliminating the interaction between the two PCPs and leaving only the C-boson interaction. **(iii)** At the *z*-axial magic interface, all 16 excited states of the CEP are in parallel jump transport from one end of the interface to the other to form a 2D boson. Obviously, unlike the bosons in elementary particles, the boson here contains all the possible *z*-space orientation states of the two electrons as well as 16 pairs of SASC electron orbitals of the two PCPs containing information on the chemical-physical structure of the molecule. This is a unique boson in the clustering of two adjacent molecules (atoms) in the mean field of condensed matter solid-state physics, and is called a clustered boson ( C-boson). This is because it always appears on the magic interface and is sometimes called a magic interface-boson or a boson trap state.

Einstein's famous observation that "quantum mechanics is incompleteness" is attributed to his deep thinking about that existing "physics is a system of logical though".57 One calls a system consisting of a finite set of initial conditions and "laws of physics" expressed in equations a computable system. In the last three decades, many scholars have undertaken theoretical explorations of incomputable physics system. Recently, J.M. Agüero Trejo et al. confirmed the existence of incomputable quantum states by "quantum random number generator" output experiments.57 In this study, interesting theoretical phenomena will once again provoke thought: is each of the 16 interfacial excited states of the CEP, which appear in sequence, a 'true' quantum state? Is each state consistent with the idea that 'all quantum states in quantum mechanics should be described in terms of "probabilities"?. Clearly, the new concept of the C-boson goes beyond the description of the behavior of the electron by the existing equations of quantum mechanics and is an incomputable quantum state in a disordered system. In fact, in the two-electron theory, neither the mass, momentum, and velocity of atoms and molecules nor the actual geometric scale of atoms and molecules (the scale is expressed only in terms of energy) is discussed; Even the time variables are represented by relaxation times from one to nine closed loops. The two-electron theory focuses only on the geometry of positive and negative charge re-equilibrium due to the emergence of "CEP interface excited states" in SASC adjacent two-molecules / two-clusters. Different soft matter molecular systems, such as foams, gels, proteins, enzymes, cell membranes, etc. have their own different de Janis $n$ = 0 emergent geometries, and we can't predict them, we can only discover them one by one!

# VI. CONCLUSUONS

The nature of glass states is the incompleteness of existing quantum mechanics. In the glass state, along the $z$-axis of the local region, each atom (molecule) has 4×16 cluster contact angle lines. On each line, there is a $\theta_\lambda$ point where two synchronous antisymmetrically coupled orbital electrons in two hard spheres meet and overlap, which cannot be handled by the one-electron theory. The second solution of $n = 0$ theory converts the 16 pairs of synchronous anti-symmetrically coupled $\theta_\lambda$-$\theta_{\lambda^*}$ points of two adjacent atoms (molecules) into 16 $z$-axial repulsive electron pairs at the two-atom (molecule) overlapping magic interface, which are the 16 interface excited states of a coupled electron pair of two adjacent atoms (molecules) and are also the electron pairing in high-temperature superconductivity.

The glass state is characterized by the parallel jump transport of the mean-field hard sphere along the 4×16 equipotential discrete points on the 4-diagonals of the cage, breaking the translational symmetry of parallel transport in topology. Thus, de Gennes's prediction was fulfilled: glass transition can be explained in simple terms: Near the absolute temperatures, the system enters a glass state when a 2D dynamic $V_0$-cluster vector in the form of a cubic lattice /equilateral hexahedron generated by four magic interfaces appeared in each local region. As the temperature increased, an increasing number of 2D $V_0$-cluster vectors in different directions are inversely cascaded along the direction of the first $V_0$-cluster vector in the local region. When the inverse cascade results in the first 2D +$z$-axis soft matrix centered on $a_0$-HSM in the reference local region, the total potential energy of the $z$-axially repulsive electron pairs at the 320 magic interfaces in the soft matrix is balanced by the random thermal vibration energy of the glass transition temperature, and the number of positively charged particles at the +$z$-axis 1/16 potential point in the soft matrix is 16 more than that at the –$z$-axis 1/16 potential point. As a result, the soft matrix obtains one degree of freedom energy that jumps along the positive $z$-axis. When the reference soft matrix disappears because of the appearance of the four surrounding soft matrices, the equilibrium positions of the positive and negative charges are at the +$z$-axis point $n_z$, causing the 200 $z$-axis hard spheres projected into the $a_0$-soft matrix to jump a small $n_z$ step along the +$z$-axis ($n_z \leq 0.036$, which is less than the molecular $z$-axial vibration scale of 0.1). The subsequent soft matrices that appear one after another in the reference region are all along the +$z$-axis, and the 2D magic interface-bosons in all these soft matrices are strongly correlated along the $z$-direction, such that the generation energy and disappearance energy of the first +$z$-axis soft matrix are shared by all subsequent +$z$-axis soft matrices in the local region. This results in a cooperative jumping pattern of all molecules in the local region being a magic interface-boson solitary wave in the $z$-direction.

# ACKNOWLEDGMENTS

The authors would like to thank Professor Wu Dacheng of Sichuan University, who promptly sent the articles of de Genève in 2002 and 2005 to the author and had useful discussions with him on $n = 0$.

# AUTHOR DECLARATIONS

### Conflict of Interest

The author declares no competing interests

# DATA AVAILABILITY

The data that support the findings of this study are available within the article and its supplementary materials.

# Supplementary materials

# Cluster model of molecule

**Jia-Lin Wu**

College of Material Science and Engineering, Donghua University, Shanghai, 201620, China. **Email**: jlwu@dhu.edu.cm

## Abstract

The term "cluster model of molecules" was coined by de Gennes. De Gennes points out that the key to the (spin) glass theory is to find the interaction that exist only between two adjacent hard sphere molecules (HSMs) in the mean-field. We now find that this interaction is a magic-interface 2D vector of two HSMs overlapping by 0.27%, containing a clustered boson and a two-electron trap state located at 0.019°, as deduced from the de Gennes $n = 0$ theory. There are six sections and seven figures in this supplementary material.

### SM 1. Why is five-HSM / five-cluster / five-local field model ?

--The first step of the model: The glassy structure is an inverse cascade pattern of nine clusters from small to large along one direction. In the solid state, molecules can only form 2D soft matrices with collective directed leaps, and *there is no so-called (independent particle) molecular dynamics.*

### SM 2. Equivalent representation of 2D cluster and interface excited states

--Each cluster must be a 2D vector, the interface of the adjacent two-molecules is a 2D magic interface-boson vector.

**SM 3. Clusters percolation transition and soft matrix**

---Percolation transition is a necessary condition to induce high-temperature superconductivity.

SM 4. **Theoretical derivation of anomalous viscosity of entangled polymer melts**

== The mode of concerted motion of the molecules in the melt is the same as in the glass transition. The difference is that in the glassy and glass transitions, each local region can generate only one direction of soft-matrix hopping solitary wave along one direction of the inverse cascade, whereas the melt can generate five soft-matrix hopping solitary waves along five different directions. The total number of soft matrix degrees of freedom in the five directions is equal to or proportional to the total number of degrees of freedom of the three soft matrix hopping solitary waves along the **z**-, *x*- and **y**-axes.

**SM 5. Cluster structure of molecules in liquid state**

---Even the liquid molecule is a soft matrix jumps instead of a molecule

**SM 6. Correlation of 4-diagonal closed-loop jump paths for all HSMs in a cluster**

---Necessary conditions for the realization of high-temperature superconductivity

**Figures S1-S7**

**Fig. S1. The glassy ideal disordered system has a stable dynamic ordered structure**:

--- Each molecule in the local region has the same cage and orientation, and the positively charged particles (PCP) of each molecule make closed-loop jumps along 4-diagonal paths on the cubic lattice (equilateral hexahedron) and synchronous inverse-symmetric coupling jumps with the neighboring PCP on each diagonal.

**Fig. S2 Four equivalent representations of 2D cluster vectors in different scenarios**

**Fig. S3 The image of the percolation transition and magic number 14**

---The key concept is that the chain of self-avoiding random walk of de Gennes $n = 0$ is a free chain in 2D space ($z$-space).

**Fig. S4 Geometric frustration of 2D soft-matrix**

-- Soft matrix gets a jumping degree of freedom in geometric frustration.

**Fig. S5 The concepts of equivalent particles and equivalent chains**

--Similar to the reciprocal lattice vector of the solid lattice, the tube model in the glassy disordered system has the concepts of equivalent particles and equivalent chains

**Fig. S6 Excluded volume of liquid hard-sphere molecules.**

**Fig. S7 Distribution of 4-diagonal closed-loop paths in a cluster**

---A prerequisite for high-temperature superconductivity

## Introduction

**1.** Chemists emphasize that the spatial geometry of a molecule is a key factor in a chemical reaction. However, no matter how complex the chemical structure of a molecule, each molecule that interacts with an infinite number of molecules is a hard spherical molecule (HSM) in the Leonard-Jones (L J) potential field. The question is, in a chemical reaction, in the collision of two HSMs, how do two electrons escape from their respective HSMs? Also, how is the exclusion volume of a molecule strictly defined? Are there interactive interfaces between two adjacent HSMs? and if so, what are the shapes of these interfaces when the HSMs move? In exploring the cluster model of the glass state, we encounter these fundamental questions in interface science. The answer is that each molecule in the mean field has, in addition to the morphology of the HSM, a 2D cubic lattice/equilateral tetra hexamer cluster morphology (also the excluded volume of this molecule) and consists of four magic interfaces formed by the sequential overlaps of four adjacent HSMs of which this HSM is respectively synchronously antisymmetrically coupled (SASC), each of which is a geometric carrier of 16 interface excited states of the coupled electron-pair of the two HSMs.

**2** There are more than a dozen existing theoretical models and views of glass and glass transition, including free volume, cage, trap, mode-coupling, random first-order transition, boson peak, Johari-Goldstein fast-slow-relaxations, heterogeneity and potential energy landscapes etc., but cluster model of molecules is missing. In 2002, de Gennes proposed a cluster model of molecules [1] in contact with various schools of thought, with the central idea of looking for the interactions that exist only between adjacent two HSMs.

**3.** The reason why it is so difficult to establish a cluster model of molecules is that every two adjacent HSMs in a cluster undergo a "mutation" in the mean-field HSM model, which changes from a one (single)-electron theoretical description to a two-electron theoretical description, that is, the sequential appearance of the 16 $z$-axis interface excited states of the coupled electron-pair of two molecules (atoms) (16 $z$-axis repulsion electron pairs spaced 5.9987° apart) makes the overlapping magic interface a 2D vector. The "mutation" of overlapping two adjacent molecules in the two-electron theory, which de Gennes calls the neighborhood effect, surpasses the "phase transition" in the one-electron theory. The neighborhood effect of constantly updating the chemical structure of adjacent two-molecules will be a new way for physics to transition to biology.

**4.** The two-electron theory is a magic-interface theory that describes the 16 $z$- azimuth interface excited states of the coupled electron pair of two neighboring HSMs. As the 16 $z$-azimuthal interface excited states appear sequentially, a new 2D vector is formed, called the C-boson of this two HSMs. A $z$-axial boson plus a $z$-axial two-electronic trap state of 0,019° form a z-axial magic interface, which is the unit vector that forms a 2D cluster in the $z$-direction, represented by an arrow on a 2D projection plane in the perpendicular $z$-direction.

**5.** In 2005, de Gennes pointed out [2] that if we want to explain the glass transitions in simple terms to our students, the way to do it is to refine the cluster contact picture of adjacent HSMs depicted in de Gennes' 2002 article entitled 'A simple picture for structural glasses'. De Gennes' picture is a cluster-contact between HSMs, which can only be the contact of two electron orbitals of the two-HSMs. Now this picture has a more perfect magic-interface update diagram. The marvelous thing about the magic interface is that the two HSMs overlap by 0.27% (any attempt to change the overlap rate does not satisfy the geometric condition of de Gennes $n = 0$) on the ±$y$-axis or ±$x$-axis, but on the $z$-axis, 16 tangent electron pairs of the two electron orbitals of the two HSMs are converted into 16 repulsive electron pairs along the $z$-axis. In the de Gennes n =

0 theory, each pair of tangent points is an eigenvalue of the interface excited state and quenched disorder of the coupled electron pair [4]. This magic interface properties will cause the existing thermodynamic statistical methods of independent particles to fail over the entire temperature range of solid-liquid transition. The new statistical method for the magic interface (confluence of thermodynamics and kinetics) consists of the following four points. **(i)** *The HSM in clustering is a vector* in the direction of the $q$-axis ($z$-axis) of the Lennard–Jones (L– J) potential, from which the concepts and symbols of $z$-component HSM-$a_0$ ($z$-$a_0$), $-z$-$c_0$... are required. **(ii)** The cage where the HSM vibrates is a cubic lattice (or equilateral hexahedron) with a side length of $2\Delta d \approx 0.1103$ (*Measuring lattice length using unit dimensionless potential well energy*, because the vibrational scale of the molecule is controlled by the equipotential plane)**.** **(iii)** Since the 2D soft matrix (i.e., the largest cluster) jumps instead of the molecules, all the soft matrices that appear in each local region over time are in the same direction at the glass transition temperature. Thus all soft matrices (excited and not yet excited) in the glass transition local region are in the same direction, and they share the generative energy $k_B T_g°$ of the first soft matrix, resulting in a cooperative hopping mode of all molecules in the local region being a magic-interface solitary wave. **(iv)** The physical image is $4 \times 16$ electron orbitals that appear sequentially filling the gaps in a 2D dynamic cubic lattice of four magic -interfaces, see **figure 3c** in [4]. Therefore, for beginners, understanding these four new concepts in disordered systems, the magic interface and the C-boson and the two-electron trap, as well as the nine 2D clusters (nine 2D interface excited loops in the inverse cascade in which all of the interface excited states of the coupled electron-pairs are orientated along the z-axis), grasps the core of the glassy state.

## SM 1. Why is five-HSM / five-cluster / five-local field model ?

**(1)** Need to build a soft matrix of the largest cluster along one direction from small to large

Molten polyester can transform entangled random macromolecules into structurally stable fully oriented polyester fibers (the glass transition has been complete) at a super-stretch rate of 30,000% within a few milliseconds under super-high-speed spinning conditions. However, under normal spinning speeds, the resulting fibers are not fully oriented and have unstable structures, and it takes hours or even days to complete the glass transition. The super-high-speed spinning line of polyester melt can link the abnormal viscosity of the entangled polymer melt to the super-tensile hydrodynamic mode and the oriented glass transition within a few milliseconds. One possible explanation is that only the largest 2D cluster (soft matrix) can move. There is a plurality of spatially oriented soft matrices in each localized region of the melt. All soft matrices in the supercooled liquid on the molten super-high speed spinning line are facing the same direction. The normal glass transition involves randomly selecting one direction of the soft matrix in each local area. In the super-high-speed spinning of polyester, within a few milliseconds, the temperature of the molten filament dropped from 300 degrees to room temperature and was stretched by more than 30,000%, which indicates that the formation of the soft matrix is independent of temperature and material deformation.

Experimental data on the cooperative orientation activation energy, $\Delta E_{co} = 2035$ ($k_B T$), of an on-line measurement of polyester under super-high-speed spinning [3] supports this view. In the experiment, $1/6\ \Delta E_{co} = 339$ ($k_B T$) is the energy of the glass transition temperature $T_g$ ($\approx 339\ K \approx 67°C$) of polyester. The coefficient 1/6 is derived from the ideal random orientation distribution of macromolecules in the melt. This may mean that the average orientation energy of a soft matrix, whether in the melt or in the glass state should be $k_B T_g$, independent of system temperature $T$.

On the other hand, whether glass state, glass transition, or moltenthe, the average (directionally coupled electron-pair interface excited state) energy of each soft matrix is Hamiltonian $H = k_B T_g°$ emerging in the local domain,, but the direction of all soft matrices in the system is random. Therefore, the average energy $H$ of the system can be expressed by the disorder energy $k_B T$ whose "molecular scale" reaches the soft matrix scale, that is, $H = k_B T_g$..

**(2)** The largest cluster (soft matrix) on the 2D $x$–$y$ projection plane is oriented along the $z$-axis

Based on (1), it can be reasonably proposed that the molten $z$-component chains of the entangled polymer "quench disorder" into a ±$z$-axis oriented glassy structure on the $x$– $y$ projection plane.

**(3)** Based on (2), in order to make all HSMs in the 2D lattice oriented along the ±$z$-axis, there must be one or more interface excitation (IE) closed loops on the four sides of each HSM to form a dynamic Ising model, see **Fig. 1** and **figures** S2, S3 and S4.

**(4)** The constructed soft matrix must contain about 200 HSMs

200-HSM is the critical molecular weight of entangled polymer chain obtained through experiments, and requires theoretical support or to verify theory. If it is considered that 200 HSMs are also 200 different spatiotemporal states of a $z$-component chain on the molten super-high-speed spinning line, and are statistically equivalent to 200 $z$-direction chain units on 200 different chains, the projection of these 200 $z$-direction chain units on the 2D $x$-$y$ plane is the largest orientation cluster in the glass state, that is, in the dynamic balance of the generation and disappearance of the soft matrix, the soft matrix must contain an average of 200 HSMs. The five-HSM/five-cluster /five-local field model can meet this requirement. The 3-HSM, 4-HSM or 6-HSM models cannot achieve this goal.

**(5)** In order for the local center $a_0$-HSM to establish three independent walking 2D soft matrices on the three projection planes in the $x$-, $y$- and $z$-axes, respectively

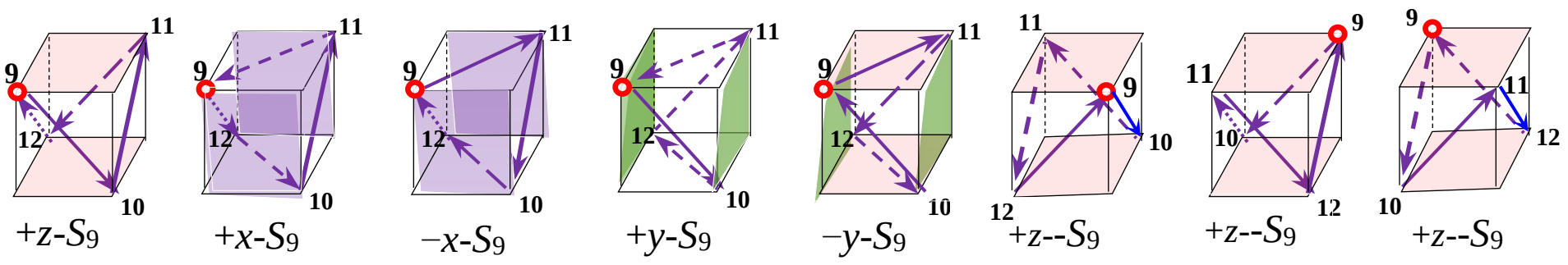

**Figure S1**. PCP ($a_0$) located at the starting point of the 4-diagonal closed loop (small red circle, point 9), selects 3 different closed loop paths among the 4 vertices (4 trap states) of its 2Δd cubic lattice to cycle 9 times forming three soft matrices, a. +$z$-$S_9$, b. +$x$-$S_9$, d. +$y$-$S_9$), which walk independently along the +$z$-, +$x$- and +$y$ axes, respectively. There are four ways to form the same +$z$-$S_9$, which are a, f, g, and h.

Based on (1), if it is considered that jumping in a certain direction in the glass transition is the largest cluster soft matrix rather than molecules, in order for the molecules to travel randomly in the melt, it must be possible to respectively construct three independently walking 2D soft matrices, ±$z$-$S_9$ ($a_0$), ±$x$-$S_9$ ($a_0$), ±$y$-$S_9$ ($a_0$), of $a_0$-HSM along the ±$z$-, ±$x$- and ±$y$-axis[14]. The formation of the +$z$-$a_0$-soft matrix is the 9 closed-loop jumps of the positively charged central particle of $a_0$-HSM [PCP($a_0$) choosing a 4-diagonal path in the +$z$-direction on the 2Δd cubic lattice / equilateral hexahedron. Only five-HSM model can do this, 3-HSM, 4-HSM or 6-HSM models cannot build a micro-cubic lattice in two-HSM clustered collisions. It also shows that there can be three directions of 2D soft matrix potential energy in 3D space, respectively, balanced with thermal disordered kinetic energy $k_B T_g$. Since the central vacancy of the soft matrix is the soliton surrounded by C-bosons, (a), (b) and (d) reveal the mechanism by which three 4-diagonal closed-loop hopping paths of PCP($a_0$) are solitons walking independently in the three directions of $z$-, $x$- and $y$-axis.

**(6)** 200-HSM soft matrix can explain geometrical frustration

Based on (4), there must be a mechanism to interrupt cluster growth. The largest cluster is also controlled by geometrical frustration. A mechanism including geometric frustration must be found to interrupt the cluster growth. Five-local field model can do this, seeing **Figure S3**.

**(7)** Can be linked to the existing concept of fluctuations in molecular number density

Each interfacial excitation (IE) closed loop in the inverse cascade corresponds to a "local phase transition" and results in an abrupt increase in the change in molecular number density (**Table 1** in the main text).

**(8)** A strict definition can be given for the excluded volume between polymer chain units (HSMs)

An excluded volume for $a_0$-HSM is the volume enclosed by a 0.27% sequential overlap of the $a_0$-HSM with its four adjacent HSMs, which is not occupied by other HSMs. That is, the excluded volume of the molecule is actually the molecular magic-interface loop (MMIL) in the form of a dynamic equilateral hexahedron, bounded by the orientation four magic interfaces.

. **(9)** Satisfy the clustered fluctuation symmetry in the random system

In the soft matrix with $a_0$ as the center, $a_0$ collides sequentially with its four adjacent molecules, $b_0$, $c_0$, $d_0$ and $e_0$, preferentially form $V_0$ ($a_0$) cluster. Then, based on the fluctuating symmetry, clusters centered on $b_0$, $c_0$, $d_0$ and $e_0$ are sequentially formed: $V_0$ ($b_0$), $V_0$ ($c_0$), $V_0$ ($d_0$) and $V_0$ ($e_0$). After that, $V_0$ ($a_0$) sequentially interacts with $V_0$ ($b_0$), $V_0$ ($c_0$), $V_0$ ($d_0$) and $V_0$ ($e_0$) to form $V_1$ ($a_0$) cluster.... In the five-local field, the fluctuation symmetry of the five soft matrices is self-similar to the fluctuation symmetry of the five clusters in the $a_0$ soft matrix, see Figure S3. The advantage is that the energy of rearranging $a_0$-soft matrix is $kT_m° = kT_g° + 4\varepsilon_0$, when $V_8$ ($b_0$), $V_8$ ($c_0$), $V_8$ ($d_0$) and $V_8$ ($e_0$) appear in fluctuations. $\varepsilon_0$ is the potential well energy.

## SM 2. Four equivalent representations of 2D cluster vectors in different scenarios

The interface excitation (IE) state is actually an overlap between the centre $a_0$ -HSM and one of its adjacent HSMs ($b_0$, $c_0$, $d_0$, and $e$) to generate a 2D magic-interface vector state, and when $a_0$ overlaps with $b_0$, $c_0$, $d_0$, and $e_0$ in turn, a $z$-axis dynamic HSM cubic lattice (HSCL) is formed, and its interface consists of four magic interfaces around $a_0$. These five HSMs are referred to as the 0th cluster +z-$V_0$ ($a_0$) of the $a_0$-HSM, and these four IE arrows are labelled as the 0th +$z$-$V_0$ ($a_0$) loop of the $a_0$ -HSM, and in de Gennes $n$ = 0 theory, the +$z$-$V_0$ ($a_0$) loop can be defined as the +$z$-axis MMIL state around $a_0$, denoted as +$z$-$S_1$($a_0$) (FIG.1a in the main text). The subscript 1 indicates that the number of closed loops around $a_0$ of the four magic interfaces is 1. +$z$-$S_4$ ($a_0$) indicates that the number of closed loops around $a_0$ of the four magic interfaces is 4, which is equivalent to a –$z$- $V_3$ ($a_0$) loop or –$z$-$V_3$ ($a_0$) cluster. That is, .$z$-$V_i$ ($a_0$) loop = $z$-$S_m$ ($a_0$), $m = i$ +1.

The four adjacent domains of the $A_0$-domain, $B_0$-, $C_0$-, $D_0$- and $E_0$-domain, all have the same 5-HSMs and 5-clusters as the $A_0$-domain, centered on the $a_0$-HSM at the center of the respective domain. For example, in the $B_0$-field along the $\mu_b$-direction, its five clusters are respectively labeled as $\mu_b$-$V_{i\ B}$ ($a_0$), $\mu$-$V_{i\ B}$ ($b_0$), $\mu$-$V_{i\ B}$ ($c_0$), $\mu$-$V_{i\ B}$ ($d_0$) and $\mu$-$V_{i\ B}$ ($e_0$). Note that there are topological connections in the cluster model: The four directions of the four clusters in the $A_0$-field, $\mu_b$-$V_0$ ($b_0$), $\mu_c$-$V_0$ ($c_0$), $\mu_d$-$V_0$ ($d_0$) and $\mu_e$-$V_0$ ($e_0$), are also the four directions of the four fields, $B_0$-, $C_0$-, $D_0$- and $E_0$-field. After averaging five soft matrices with different orientations in 5-local field ($A_0$, $B_0$, $C_0$, $D_0$ and $E_0$) centered at $A_0$ of $a_0$, the $z$-axial $a_0$-soft matrix will contain 196

+ 4 $\sigma$-HSMs and five HSCLs "central cavity spaces". The 4 $\sigma$-HSMs here are the four $\sigma$-HSNs in $V_{0A}$ ($a_1$) of $a_1$ on the $z$-component chain $N_z$ in the SM 3.

In a referenced $A_0$-domain (-field, SM 2), by replica symmetry, $\mu_b$-$V_0$ ($b_0$), $\mu_c$-$V_0$ ($c_0$), $\mu_d$-$V_0$ ($d_0$) and $\mu_e$-$V_0$ ($e_0$) in four different directions adjacent to $a_0$ appear in sequence, and project again in sequence to the $z$-axial and clustered collide-overlap with $z$-$V_0$ ($a_0$) to form a new 2D $-z$-$V_1$ ($a_0$)-cluster [and $-z$-$V_1$ ($a_0$)-loop] and L– J potential $f_1$ ($\sigma_1/q_1$), and $\sigma_1$ ($a_0$) has 17 $\sigma$ HSMs (SM 3). And so on, along the $q$-axis of the nine $z$-axial L– J potentials, nine 2D $a_0$-clusters $V_i$ ($a_0$) cluster, and nine 3D $a_0$ hard-spheres $\sigma_i$ ($a_0$) inverse cascades from small to large, up to the largest 2D $z$-$V_8$ ($a_0$)-cluster (soft matrix).

Thus, the four equivalent representations of a 2D cluster vector centred on $a_0$ are $S_m$ ($a_0$)**;** $V_i$ ($a_0$) cluster**;** $V_i$ ($a_0$) IE loop and $V_0$ ($a_0$, $\tau_i$), the last of which appears in the magic interface-boson solitary wave of the glass transition, $V_0$ ($a_0$, $\tau_i$) denotes that the relaxation time of the four magic interfaces around $a_0$ in $V_0$ ($a_0$) is $\tau_i$, i.e., the four magic interfaces around $a_0$ complete the $i$-th closed loop, which is also equivalent to the $V_i$ ($a_0$) cluster.

## SM 3. Clusters and percolation transition and soft matrix

The clustering constant for HSMs: Δd = 0.05514…. The "cage" of the HSM is a micro-cubic lattice (equilateral hexahedron) with a side length of 2Δd, and the HSM makes 1 to 9 closed-loop jumps along the 4×16 discrete points on the micro-cubic lattice. Each face of the 2Δd micro-cubic lattice in the cluster is an equipotential face, which has a potential energy of 1/16 relative to the center of the lattice (let the potential energy at the bottom of the potential well be 0 point). When the PCP of $a_0$-HSM starts from point 9 in Fig. 2 along the 4-diagonal closed loop path on the 2Δd cubic lattice and returns to point 9, the $z$-$a_0$-HSM obtains the 1/16 potential energy on the +$z$-axis, and the $-z$-$c_0$-HSM that circulates synchronously with $a_0$ obtains the –1/16 potential energy on the $-z$-axis. This means that the magic-interface and the C-boson and trap states are located at point 0 of the potential energy of the positive-negative charge equilibrium, i.e., at the bottom of the potential well (the nature of the magic-interface). If each HSM in the $V_i$ loop is along the +z-axis (labeled +$z$-HSCL), then the potential energy on the +$z$-axis is 1/16, and if its IE loop is along the $-z$-axis (labeled $-z$-HSCL), the potential energy on the $-z$-axis is –1/16.. In Figure. S2, (a) and (b) shows the charge balance between $V_2$ ($a_0$) and $V_2$ ($b_0$) in the cluster. Dynamic charge balance occurs between 5 +$z$-HSCLs in $V_2$ ($a_0$) and 5 $-z$-HSCLs in $V_2$ ($b_b$). This property applies to the clustering of all two adjacent clusters.

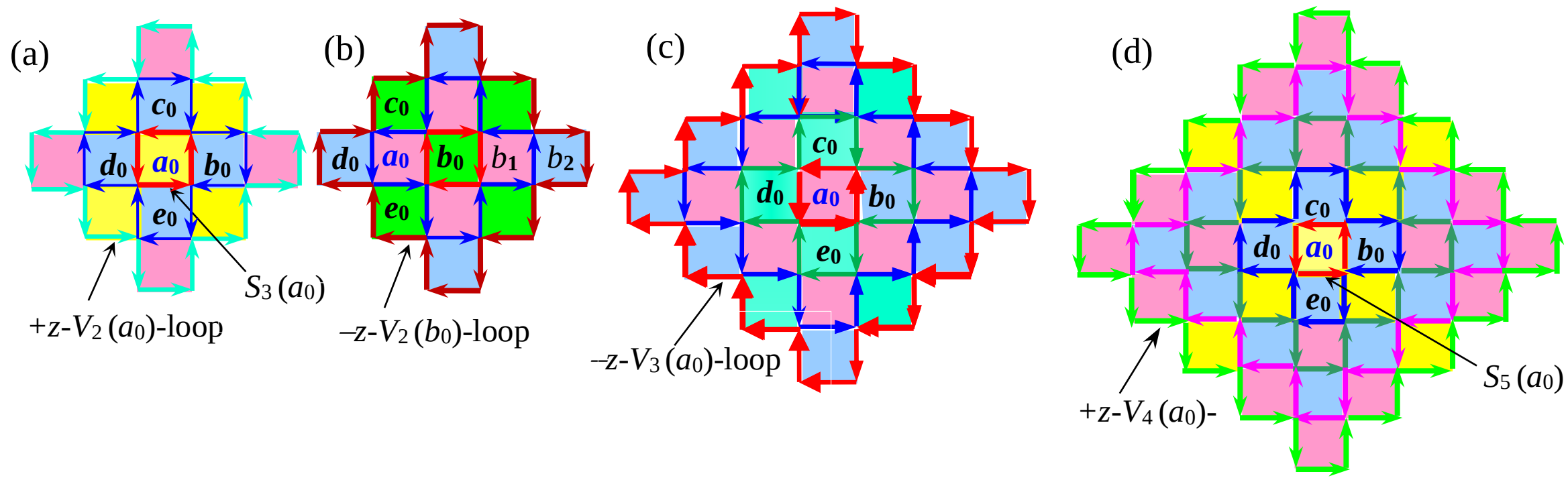

**Figure S2**. Mosaic structure of positive-negative charges in cluster enlargement. **a**. $V_2$ ($a_0$) cluster contains 33 HSMs, 20 of which are edge particles. There are 9 +$z$-HSNs (red and yellow) and 4 $-z$-HSNs (blue) in the $V_2$($a_0$)-loop. After the energy on the +$z$-axis and $-z$-axis cancel each other out, there are 5 +$z$-HSNs left (yellow). **b**. There are 5 $-z$-HSNs left (green) in $-z$-$V_2$ ($b_0$)-loop. **c**. $V_3$ ($a_0$) cluster contains 53 HSMs, 28 of which are edge particles. There are 9 +$z$-HSNs (red) and 16 $-z$-HSNs (blue and lake

green) in the $V_3$ ($a_0$)-loop, and there are 7 $-z$-HSNs left (lake green). **d**. $V_4$ ($a_0$) cluster contains 77 HSMs, 36 of which are edge particles. There are 25 $+z$-HSNs (red and yellow) and 16 $-z$-HSNs (blue) in the $V_4$ ($a_0$)-loop, and there are 9 $+z$-HSNs left (yellow).

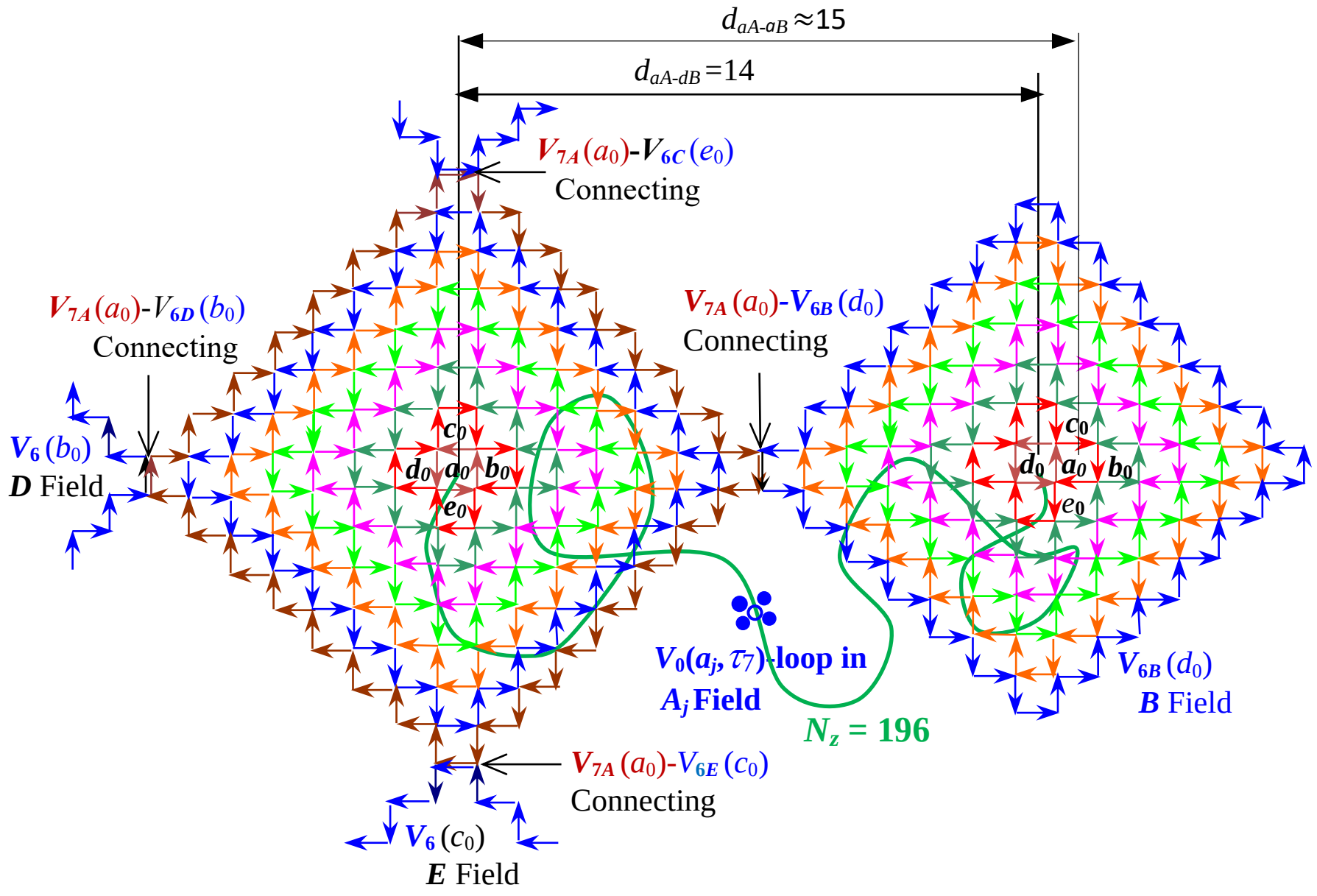

**Figure S3 The image of the percolation transition connected by $V_{7A}$ ($a_0$)-loop and $V_{6B}$ ($d_0$)-loop**. When $V_{7A}$($a_0$)-loop appears, the $V_{7A}$ ($a_0$)-loop formed by 60 IE arrows is connected to the $V_{6B}$ ($d_0$)-loop formed by 52 IE arrows. Two of the IE arrows merge into one arrow at the connected interface and reduce the two edge particles between the two clusters $V_{7A}$ ($a_0$) and $V_{6B}$ ($d_0$), thereby reducing the lattice Hamiltonian. $V_7$ ($a_0$) cluster contains 173 HSMs, 60 of which are edge particles. There are 64 $-z$-HSCLs and 49 $+z$-HSCLs in the $V_7$ ($a_0$)-loop, and there are 15 $-z$-HSMs left. $V_6$ ($a_0$) cluster contains 137 HSMs, 52 of which are edge particles. There are 49 $+z$-HSCLs and 36 $-z$-HSCLs in the $V_6$ ($a_0$)-loop, and there are 13 $+z$-HSCLs left. The critical length of polymer chain entanglement is 200 HSMs. The 196 HSMs (green curve) in the *A*-field are represented as $a_0$, $a_1$, $a_2$,... $a_j$,... and $a_{195}$, and the field of the $a_j$-th HSM forming the $a_j$-soft matrix is denoted as $A_j$-field. The other 4 HSMs (blue) out of the 200 HSMs are the four HSMs that form +z-$V_0$ ($a_j$, $\tau_i$), here +z-$V_0$($a_j$, $\tau_i$) = +z-$S_m$ ($a_j$), $\tau_i$ is the *i*-th relaxation time, which means that the four magic-interfaces around $a_j$ have completed $m$ ($m$= $i$+1) closed loops. Since the background temperature is $T_k$, +z-$V_0$ ($a_j$, $\tau_i$) is now +z-$V_0$($a_j$, $\tau_7$). The *z*-component chain $N_z$ is a random chain in *z*-space (2D *x*-*y* projection plane) with one end $a_0$ is located at $a_0$ in the $A_0$-field (domain), and the other end $a_{195}$ is located at $d_0$ in the $B_0$-field. Due to random distribution, $196 = 14^2 = (d_{aA-dB})^2$, $d_{aA-dB}$ is the mean square end distance of random statistics chain. Note: For the chain $N_z$ of a specific local area in the solid state, the distance between the two endpoints of the chain $N_z$ can be a random number, such as 12 or 17, etc., but for the statistical average of an infinite number of local regions, the statistical chain $N_z$ is equivalent to an ideal free chain with a mean squared end-to-end distance of 14. 14 is a de Gennes magic number [4].

**Figure S4** is an image of spin glass envisioned by Edwards and Anderson back in 1975 [5]: to construct a simple model and incorporate the two physical ingredients of geometric frustration and quenched disorder into the lattice Hamiltonian and Ising model. As you can see, the key point ofhis diagram is that the largest 2D cluster (soft matrix) is made up of 320 2D magic interfaces, . which is the largest dynamically ordered structure in the glass state.

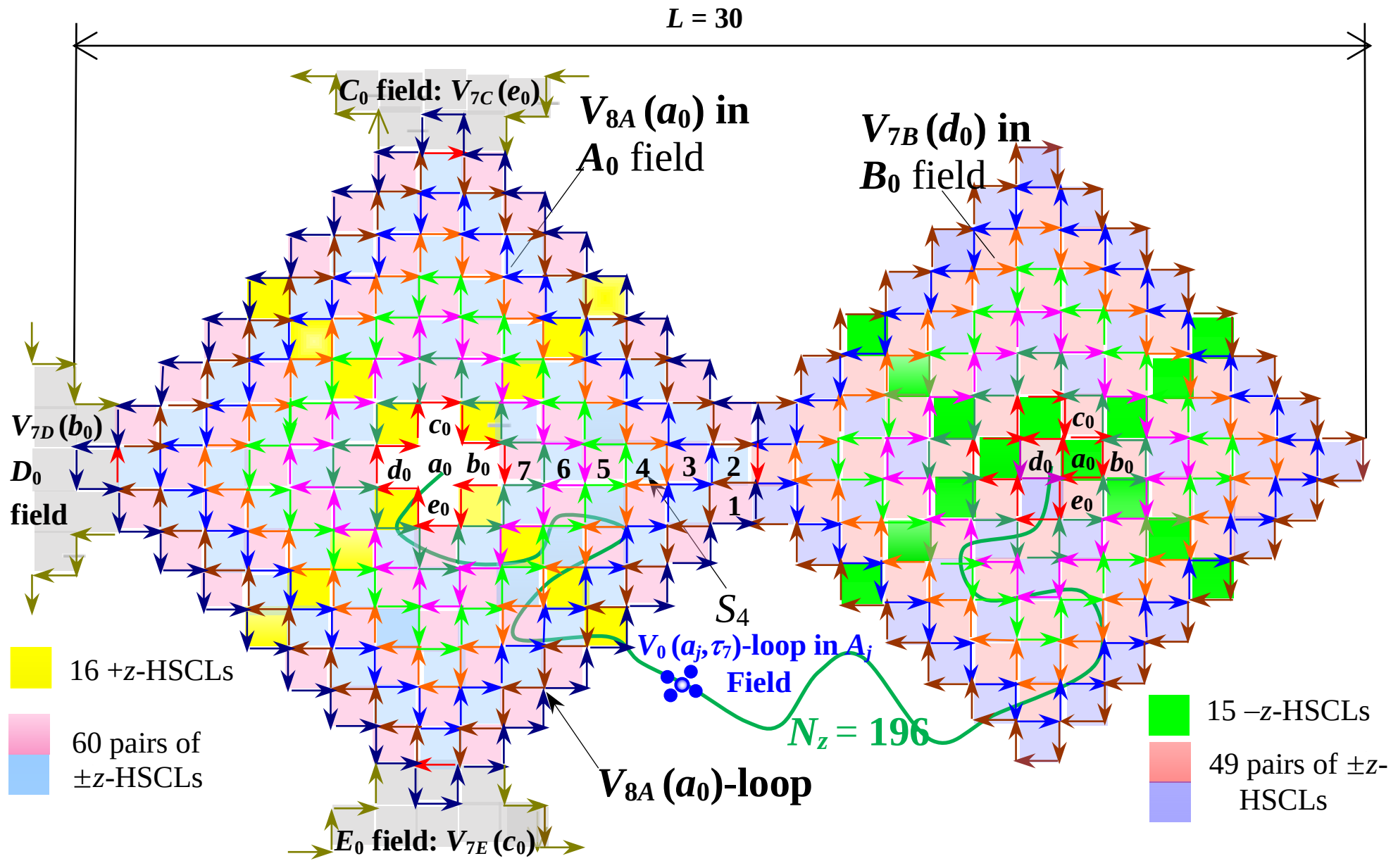

**Figure S4. The image of the $V_{8A}(a_0)$-soft matrix-loop and $V_{7B}(d_0)$-cluster-loop.** The lattice Hamiltonian to generate the soft matrix is the average energy of 60 IEs (56 dark blue arrows plus 4 reverse red arrows) with a relaxation time of $\tau_8$. When four clusters $V_{8A}(b_0)$, $V_{8A}(c_0)$, $V_{8A}(d_0)$ and $V_{8A}(e_0)$ appear sequentially, $V_{8A}(a_0)$ disappears, and the four sides (four arrows) of the MMIL of $a_0$ disappear and the vacancy space of 5-HSM appears. $V_8(a_0)$ cluster contains 200 HSMs, 60 of which are edge particles. There are 136 HSCLs, 76 +$z$-HSCLs and 60 −$z$-HSMs in the $V_8(a_0)$-loop. When the $V_{8A}(a_0)$ soft matrix appears, there are unbalanced 16 +$z$-HSCLs (yellow) in the $V_{8A}(a_0)$-IE-loop. Each +$z$-HSCL has 1/16 potential energy, and 16 +$z$-HSCLs in the $V_8(a_0)$-loop. At this time, the average position of all negative charges, the midpoint position of all $z$-axis repulsive electron pairs, is at the bottom of the potential well (set the potential to 0); And the average position of all positive charges is the equipotential point of 1/16 of the start and end points of each PCP in the 16 yellow +$z$-axis HSCLs in the soft matrix along its 4-diagonal closed-loop jump. At the glass transition temperature $T_g$, the +$z$-$V_0(a_j, \tau_7)$ of the $A_j$-field can now also be in the +$z$-$V_0(a_j, \tau_8)$-soft matrix state in the mean field representation. That is, the statistical chain of (196 + 4) HSMs in the $z$-space is a self-avoiding random walk ($n_z$-steps) chain satisfying de Gennes $n$ = 0, and the satisfying way is to sequentially appear a soft matrix composed of (196+4) HSMs,

## SM 4. Theoretical derivation of anomalous viscosity of entangled polymer melts

The rough original proof of this section is shown in Ref.12, which is now partially corrected and more accurate. Taking inspiration from SM 1 (1), we can derive the theory of glass state from the theoretical derivation of the abnormal viscosity of the 3.4 power law of entangled polymer melts. The author's research method differs from that of most scholars. First, he correctly guessed the expression for the anomalous viscosity that matched the experiment very well. For flexible chains, equation (1) derives $\eta \sim N^{3.4}$, and for non-flexible chains, equation (1) also matches the existing experimental data

$$\eta \sim N^{9(1-Tg/Tm)} \quad (1)$$

Due to the exponential relationship, the reliability of equation (1), which is highly consistent with the experiment, is high. The pattern of molecular collective jumping and the theory of glass state should be included in equation (1) , so the theoretical proof of equation (1) is an effective way to explore the nature of glass state.

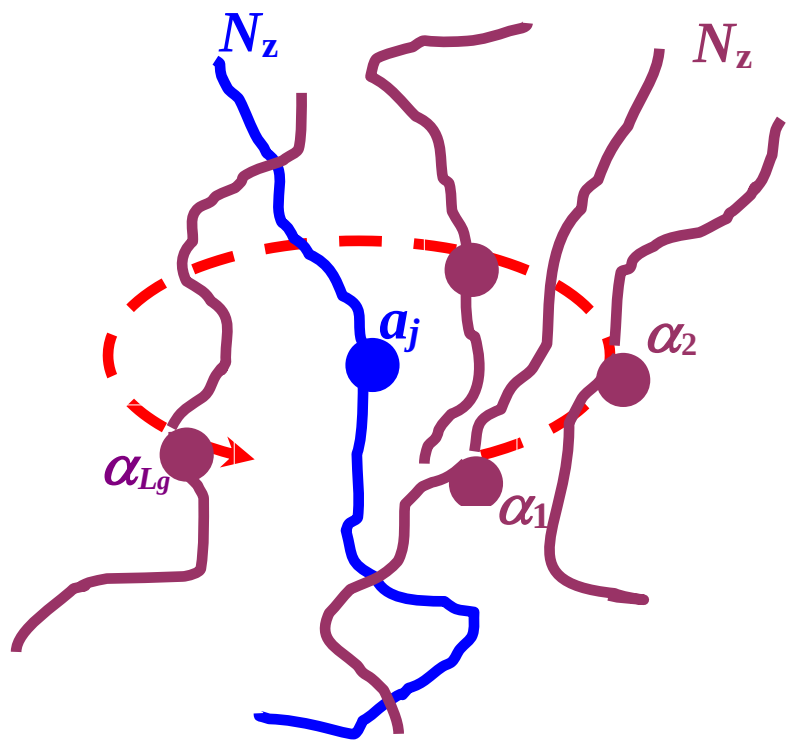

**Figure S5.** The $N$ $z$-axial soft matrices appearing in sequence share the generation energy $k_BT_g°$ of a $z$-axis soft matrix.. Due to the inverse cascading, the energies of the CEP IE states in the $a_j$-soft matrix all evolve into the $V_8$-IE-loop energy $k_BT_g°$ of the soft matrix. And $V_8$-IE-loop energy $k_BT_g°$ can be equivalent to a "loop-flow" composed of $\alpha_{Lg}$ (= $k_BT_g°$ / $\varepsilon_0$) equivalent particles ($\alpha_1$, $\alpha_2$, $\alpha_3$ ... $\alpha_{Lg}$).

De Genre's snake model gives an expression for the anomalous viscosity of the entangled polymer melt

$$\eta \sim N^3 \tag{2}$$

One possible explanation for the ultra-high-speed spinning mechanism is that the only soft matrices in all local regions in the melt that can beat are in the $z$-axial direction. Because the hydrodynamic mode is characterized by the total number of degrees of freedom (DoF) of the system. Thus, the chain length $N$ in equation (2) is replaced by the total number of DoF $N^*$ required for $N$ chain particles to spread

$$N^* = N_z^* \cdot N_x^* \cdot N_y^* = (N_z^*)^3 \tag{3}$$

Equation (2) is rewritten as

$$\eta \sim (N_z^*)^9 \tag{4}$$

The glass state theory to be sought is thus contained in the rigorous proof of equations (3) and (5).

$$N_z^* \sim N^{(1-Tg\,/\,Tm)} \tag{5}$$

In the glass transition, all $N$ soft matrices that are excited and to be excited in each local area jump $n_z$ steps along the same direction (as in the $z$-axis) in sequence. The $z$-axis repulsive electron pairs in the $N$ $V_8$-IE-loops form a dynamic "wall" of length $N$ whose energy, expressed in terms of temperature $T$, is the same as the wall energy of $a_0$-$V_8$-loop, both are $k_BT_g$.

Each equivalent particle has a unit DoF energy $\varepsilon_0$ and is located on its own equivalent chain of chain length $N_z$ ($z$-component chain of chain $N$). The probability that a $V_8(a_j)$-loop of $a_j$ occupies the $N$ associated $V_8$-loops on chain $N_z$ is $1/N$. Since the energy of $N$ related $V_8$-loops on chain $N_z$ is still numerically $k_BT_g$ (= $k_BT_g°$), the $L_g$ equivalent particle can be equivalent to the association of $N$ $V_8$-loops, as long as each equivalent particle is regarded as having a $V_8(a_j)$-loop and the probability of occupying $N$ $V_8$-loops is also $1 / N$.

Statistically, the occurrence probability $\hat{p}_+$ of the $V_8$ ($a_0$)-soft matrix is equal to the probability of simultaneous occurrence of $L_g$ equivalent particles in the $V_8$ ($a_0$)-loop:

$$\hat{p}_+ = (1/N_z)^{k\mathrm{B}Tg/\varepsilon 0} \tag{6}$$

Like the generation energy of $V_8(a_0)$-soft matrix, the O($a_0$) cavity in its center is also shared by all chain HSMs on the $N_z$ chain. Let the occupied share of $a_0$ in the O($a_0$) cavity be $n_z$, similar to Eq. (6), the probability $\hat{p}_-$ of the disappearance of $V_8(a_0)$-soft matrix is

$$\hat{p}_- = (n_z)^{k\mathrm{B}Tm/\varepsilon 0} \tag{7}$$

From $\hat{p}_+ = \hat{p}_-$ and $(1/N_z)^{Tg} = (n_z)^{Tm}$

$$n_z = (N_z)^{-Tg/Tm} \tag{8}$$

For the flexible chains $T_g/T_m = 5/8$, take $N = 200$ in Eq. (8) to adapt to the small molecule system, and obtain

$$n_z \leq 0.036 \tag{8}$$

$n_z$ is also the step-size and the number of DoF for $a_0$ to walk. Therefore, the number of DoF $N_z^*$ required to "move the entire $n_z$ steps" of the chain $N_z$ along the +$z$ axis is

$$N_z^* = N \cdot n_z = (N)^{(1-Tg/Tm)} \tag{10}$$

This is Eq. (5). Equation (10) shows that the $z$-axis IE energy in all $N$ $z$-axis $V_8$-loops in each local region, from the glass state to the molten state, is correlated with the loss of potential energy in the A-$q_L$ interval (Fig. 2) in the $z$-axis L– J potential, and maintains the value $k_BT_g^*$ (= $k_BT_g$) shared by the $N$ soft matrices that appear sequentially. In particular, at the glass transition temperature $T_g$, on average, only one soft matrix is excited in each local region, and the energy of the disappearing soft matrix is lacking, so there is almost no isolated wave beating of the soft matrix. Further, the lower the temperature is $T_g$, the less soft matrices describe the collective beating of molecules, and the greater the viscosity.

When the temperature continues to increase, the positively charged central particle of $a_0$-HSM, PCP($a_0$), jumps 9 closed loops along the clustered path of 9→12→10→11→9 in figure S1 (b) to obtain the $x$-axial soft matrix of $a_0$, written as $x$-$V_8$ ($a_0$). In order for the $x$-$V_8$ ($a_0$) to jump freely along $x$-direction, $a_0$ must consume $N_z^*$ more DoF. Similarly, in order for $x$-$V_8$ ($a_1$) to jump freely in the $x$-direction, $a_1$ must also consume more $N_z^*$ DoF. Thus, in order for $N$ $x$-axis soft matrices to jump freely in the $x$-direction, the chain $N$ seems to consume $N \cdot N_z^*$ DoF. However, as with the $N_z$ component chain, the generation of the $N$ $x$-axis soft matrix in the $N_x$ component chain is also associated with the disappearance of the A-$q_L$ interval potential of the L– J potential in the $x$-axial, and the number of DoF required for the $N_x$ chain to freely jump $n_x$ steps in the $x$-axial is $N_x^*$, that is, the number of DoF required for the $N$-chain to spread freely in $z$-$x$ 2D space is $N_x^* \cdot N_x^*$. This leads to the number of DoF required for a chain of length $N$ to spread freely in 3D space, $N^* = N_x^* \cdot N_x^* \cdot N_y^*$. This is Eq. (3), which proves that the pattern of motion of the entangled chains in the melt is that of independent walking solitons.

## SM 5. Cluster structure of molecules in liquid state

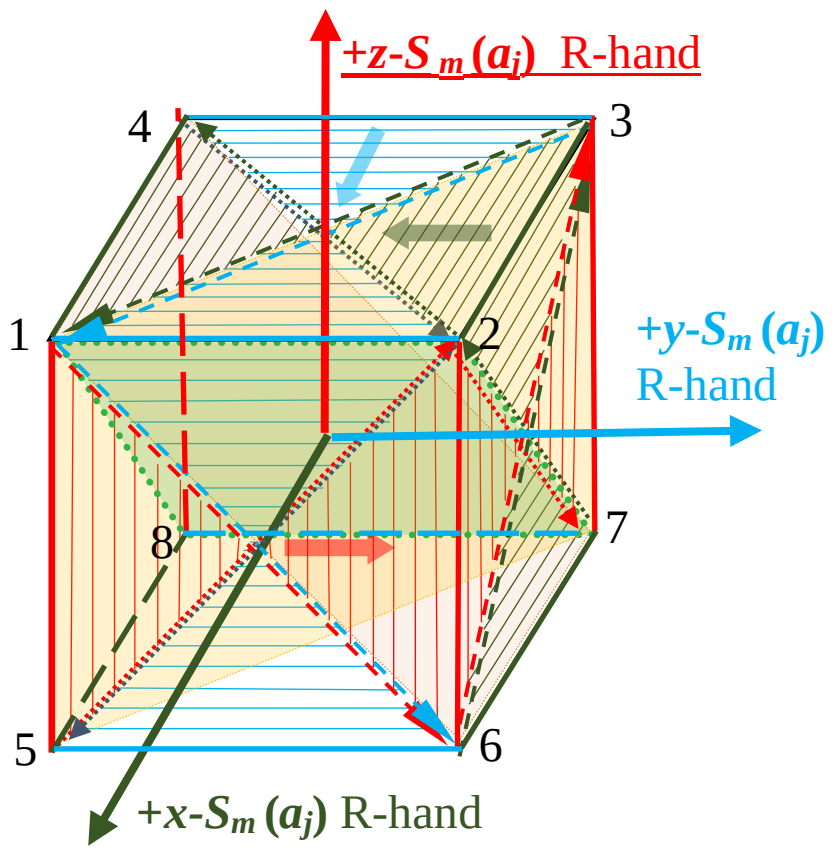

**Figure S6. Exclusion volume diagram of mean-field hard-sphere molecules in melt (liquid).** Each of the 2D MMIL can have both left and right hands, indicated by the direction of the arrow of the relative axis. In the liquid state, in addition to the various crystal structures of molecules described by quantum mechanics, there can be various dynamic equilateral hexahedral space-time structures composed of changes in quenching disordered eigenvalues caused by changes in temperature, pressure, and impurities. The 5-HSM central HSM in the liquid state has 6 ×16 spatial angle-linel states, which can form three axial 2D MMILs.

## SM 6. Correlation of closed-loop hop paths for all HSMs in the soft matrix

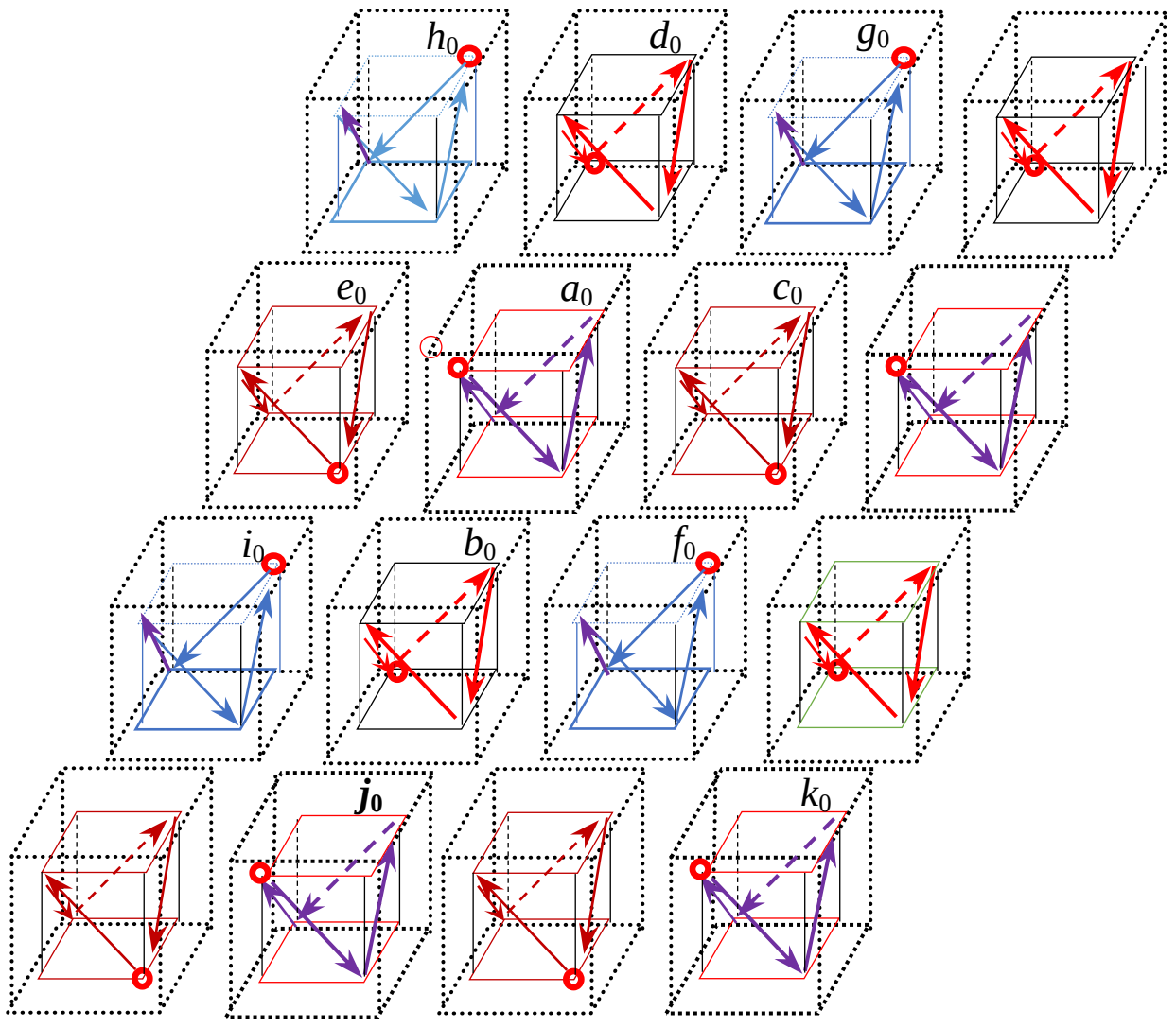

**Figure S7. Distribution of four 4-diagonal closed-loop paths (four colors) in a cluster.** Each PCP performs a 4-diagonal closed-loop jump on its 2Δd micro-cubic lattice and is marked with a small red circle at the starting position (phase) in the soft matrix. Once the PCP ($a_0$) has determined its 4-diagonal path, the 4 diagonal closed-loop paths of all in the soft matrix have been confirmed. A phase difference of $2\pi i$ is allowed between two PCPs with the same starting position, and $i$ is the $i$ in the $Vi$ cluster where the is located. For example, $j_0$ belongs to the $V_1$ cluster and its starting phase lags behind $a_0$ ($V_0$ cluster) $2\pi$, and $k_0$ belongs to the $V_2$ cluster, its phase lags behind $a_0$ $4\pi$. Each HSM in the polymer soft matrix has 3 concentric HSCLs, Figure S10. The dotted cubic lattice in the figure is a cubic lattice of $H^+$-P (the positively charged particle of hydrogen atom) that jumps synchronously with PCP.

**Figures S6** and **S7**show only the spatial geometry of the C-boson interaction states of HSM. When considering the temporal nature of an C-boson (magic-interface) that can be repeated at the same spatial location, an +*z*-axial $a_0$-soft matrix centered on $a_0$-HSM has four equivalent representations in different scenarios. They are: +*z*-$V_8$ ($a_0$)-cluster, +*z*-$V_8$ ($a_0$)-loop, +*z*-$S_9$($a_0$) and +*z*-$V_0$ ($a_0$, $\tau_8$). The fourth representation refers to the completion of 9 closed loops by four C-bosons at the four interfaces of the +*z*-axis of the (1 + $d_L$) cubic lattice of $a_0$-HSM, with a relaxation time of $\tau_8$